\documentclass[journal,transmag,onecolumn]{IEEEtran}
\usepackage{lineno}
\usepackage{color}
\usepackage{enumerate}
\usepackage{amsmath}
\usepackage{amssymb}
\usepackage{multicol}
\usepackage{graphicx}
\usepackage{array}
\usepackage{etoolbox}
\newcounter{magicrownumbers}
\newcommand \rownumber {\stepcounter{magicrownumbers} \arabic{magicrownumbers}}
\modulolinenumbers[5]

\newtheorem{theorem}{\hspace{0em}Theorem}

\newtheorem{proposition}{\hspace{0em}Proposition}
\newtheorem{remark}{\hspace{0em}Remark}

\ifCLASSINFOpdf
\else
\fi
\begin{document}
	%
	\title{Distributed Joint Sensor Registration and Multitarget Tracking \\ Via Sensor Network}

	
	\author{\IEEEauthorblockN{Lin Gao,
			Giorgio Battistelli,
			Luigi Chisci$^*$, and
			Ping Wei
			\thanks{Lin Gao, Giorgio Battistelli and Luigi Chisci are with Dipartimento di Ingegneria dell'Informazione (DINFO),  Universit\`{a} degli Studi di Firenze, Italy.}
			\thanks{Ping Wei is with the School of Electronic Engineering, Center for Cyber Security,  University of Electronic Science and Technology of China, Chengdu, P. R. China.}
		}
		\thanks{
			The work described in this paper was partially supported by the Fundamental Research Funds for the Central Universities ZYGX2016Z005 and ZYGX2016J218.}}

	\markboth{L. Gao, G. Battistelli, L. Chisci and P. Wei, Submitted to IEEE Transactions on Aerospace and Electronic Systems.}
	{}
	%



	\IEEEtitleabstractindextext{%
		\begin{abstract}
			This paper addresses distributed registration of a sensor network for multitarget tracking.
			Each sensor gets measurements of the target position in a local coordinate frame, having no knowledge about the relative positions (referred to as drift parameters) 
			and azimuths (referred to as orientation parameters) of its neighboring nodes.
			The multitarget set is modeled as an \textit{independent and identically distributed} (i.i.d.) cluster \textit{random finite set} (RFS),
			and a consensus \textit{cardinality probability hypothesis density} (CPHD) filter is run over the network 
			to recursively compute in each node the posterior RFS density.
			Then a suitable cost function, 
			expressing the discrepancy between the local posteriors in terms of averaged Kullback-Leibler divergence, 
			is minimized with respect to the drift and orientation parameters for sensor registration purposes.
			In this way, a computationally feasible optimization approach for joint sensor registraton and multitarget tracking is devised. 
			Finally, the effectiveness of the proposed approach is demonstrated through simulation experiments on both tree networks and networks with cycles, as well as with both linear and nonlinear sensors.
		\end{abstract}
		
		}

	\maketitle

	\IEEEdisplaynontitleabstractindextext

	%
	\IEEEpeerreviewmaketitle

	\section{Introduction}
	
	\emph{Distributed multitarget tracking} (DMT) on a sensor network made up of low cost 
	and low energy consumption sensors has attracted great interest
	due to the rapid advances of wireless sensor technology
	and its wide potential application in both civil and defense fields. 
	The use of such sensor networks can clearly enhance performance while decreasing cost and facilitating deployment of surveillance systems. 
	The goal of DMT is to achieve scalability and comparable performance with respect to  centralised architectures.
	Exploiting random finite set (RFS) theory \cite{mahler2007statistical,mahler2014advances}, 
	generalized covariance intersection (GCI) \cite{mahler2000optimal} and consensus \cite{olfati2007consensus}-\nocite{kamal2013information}\cite{battistelli2015consensus},
	several effective DMT approaches have been proposed \cite{battistelli2013consensus}-\nocite{uney2013distributed,guldogan2014consensus,wang2017distributed,yi2016distributed,fantacci2015consensus,suqi2018}\cite{suqi2019}
	assuming that
	all sensor nodes in the network had been correctly registered/aligned in a common global reference frame.
	
	In many practical scenarios, however, 
	the problem of sensor registration has to be tackled jointly with target tracking, 
	since, in certain circumstances, 
	it is hard to get accurate knowledge about the positions and/or orientations of the deployed sensor nodes.
	Most of the existing work on sensor registration is based on two approaches.
	In the first approach, called \emph{cooperative localization}, 
	each sensor is provided with direct measurements relative to positions of its neighbors \cite{moses2003self}-\nocite{patwari2005locating,ihler2005nonparametric,frampton2006acoustic,meyer2012simultaneous,sun2012accurate,shao2014efficient}\cite{morral2016distributed}. 
	Conversely, the second approach is based on exploiting some reference nodes of known positions (also called anchors)
	in the global coordinate system \cite{khan2009distributed}-\nocite{vemula2009sensor,cevher2006acoustic}\cite{chen2011sequential}.
	The locations of anchors are assumed known a priori or 
	can be obtained by using global localization technology such as, e.g., GPS (Global Positioning System).
	Unfortunately, however, both approaches have their limitations.
	The former requires additional sensing devices 
	for measuring the positions of the neighboring nodes, 
	and it is hard to obtain the inter-node measurements in some situations, 
	e.g. confined environments with multipath.
	The latter can only be used in some specific scenarios 
	where either prior knowledge of the surveillance area is available or 
	signals from the global localization equipment can be received.
	Conversely, in some specific applications, e.g., underwater or indoor environments,
	wherein the GPS signal cannot be received,
	this approach is not viable.
	In this paper, the interest is for a technique that neither needs sensing the positions of neighbors 
	nor the presence of reference nodes.
	
	In this respect, some interesting techniques have been recently introduced \cite{kantas2012distributed,jiang2016sensor}. 
	In particular, \cite{kantas2012distributed} exploits online distributed \textit{maximum likelihood }(ML) and \textit{expectation maximization} (EM) methods. 
	The nodes iteratively exchange the local likelihoods based on the message passing (belief propagation) technique. 
	In this approach, at each sampling interval several iterations must be carried out in order to exchange the data through the network. 
	The employed message passing method is well suited for networks with tree topology but cannot guarantee to avoid double counting in networks with cycles,
	and it is not robust with respect to changes of the network topology.
	The work in \cite{jiang2016sensor} adopted the same strategy for sensor registration as in \cite{kantas2012distributed},
	while employing consensus instead of belief propagation for message passing, thus allowing to cope with networks having cycles and time-varying topology.
	Both contributions considered sensor registration for single-target tracking under the ideal condition
	wherein the target is assumed to always exist throughout the whole observation period,
	sensor nodes detect the target with unit probability, and the sensing process is not affected by false alarms (clutter).
	
	In this paper, the aim is to solve sensor registration in the multitarget case,
	where phenomena like target existence/disappearance, missed/false detections of targets, and uncertain data associations must be accounted for.
	To the best of our knowledge, the only existing contribution in this context is the Bayesian approach of \cite{uney2016cooperative},  wherein
	a Monte Carlo method is adopted to represent and compute in each node the posterior distribution of the relative positions (drift parameters) of its neighbors.
	In \cite{uney2016cooperative},  distributed computation was accomplished by the message passing strategy 
	which can suffer from the same problems of \cite{kantas2012distributed}  with networks that change in time and/or contain loops.
	Conversely, the approach to this paper jointly solves the sensor registration and multitarget tracking problems over a sensor network
	in a distributed way by exploiting consensus. Further, estimation of both relative positions and orientations is addressed.
	
	Multiple targets are modeled as an \textit{i.i.d. cluster} RFS \cite{mahler2007phd},
	whose cardinality (the number of elements in the RFS) and the states of each set-member (target) are time-varying.
	A \textit{cardinality probability hypothesis density} (CPHD) filter \cite{vo2007analytic} can be run in each node of the network to update the local posterior density of the target set 
	with the multitarget motion model and the available local measurements.
	When the sensor relative positions and orientations are known, the consensus method can be exploited
	as in \cite{battistelli2013consensus} to fuse the local posteriors into a global one in a fully distributed fashion.
	The resulting method, referred to as \textit{consensus CPHD} (CCPHD) filter in \cite{battistelli2013consensus}, 
	provides an effective solution to DMT over a registered sensor network. 
	The fusion rule adopted in \cite{battistelli2013consensus} relies on the information-theoretic paradigm of the
	\textit{weighted Kullback-Leibler average} (WKLA) according to which the fused density is chosen as the one minimizing a special cost, defined as a \textit{weighted average Kullback-Leibler divergence} (WAKLD) from the local posteriors.
	In \cite{battistelli2015distributed} it has been proved that 
	the resulting WKLA multiagent fusion, also known as Generalized Covariance Intersection (GCI), turns out to be immune to double counting of information and is, therefore, resilient to the presence of loops in the sensor network.
	The minimum cost associated to the WKLA-fused density is known in the literature as {\em GCI divergence} \cite{suqi2018,suqi2019}. The
	GCI divergence provides a sensible measure of the degree of dissimilarity among the set of local posteriors (see \cite{suqi2018}), and can therefore be minimized with respect to the unknown drift and/or orientation parameters for sensor registration purposes. 
	
	Following the above arguments, the GCI divergence 
	is adopted in this paper as  {\it instantaneous cost} (IC) to quantify the amount of registration errors at each fusion step.
	Since the minimization of the IC would make the resulting estimates of the registration parameters sensitive to transient errors, a {\it total cost} (TC), defined as the summation of ICs 
	over fusion steps, would be a more appropriate candidate for sensor registration.
	It is shown that in the special case in which the orientation parameters are known a priori,
	the TC can be recursively computed  over time so that its direct optimization with respect to drift parameters can be a computationally feasible approach for estimating them.
	Conversely, in the case wherein both drift and orientation parameters are unknown, 
	the recursive computation of the TC is no longer possible so that direct optimization of the TC  would require excessive computational and memory loads not feasible for
	low cost sensor nodes.
	Hence, a suboptimal approach is proposed by minimizing the IC at each fusion step and then combining the resulting istantaneous estimates of the registration parameters
	according to a suitable multi-hypothesis method in order to obtain estimates that are less sensitive to transient errors.
	
	The remarkable features of the proposed sensor registration algorithm are that:
	(1) it requires no additional hardware devices, on-board and/or in the environment, for sensor localization;
	(2) it introduces no additional data exchanges, 
	only slight extra computational load 
	and memory space as compared to the original CCPHD filter.
	Further, the proposed algorithm is insensitive to the type of sensor network, a feature that is inherited from the properties of WKLA fusion and consensus \cite{battistelli2015distributed}.
	Since i.i.d. cluster processes represent a quite general family of RFS processes, 
	and they can also be used to approximate the majority of labelled and unlabelled RFS processes  \cite{mahler2014advances,vo2013labeled,tutorial}, 
	the proposed sensor registration algorithm can be flexibly combined with any DMT algorithm.
	
	{
		It should be noted that,
		in our recent work \cite{gao2019distributed},
		a similar approach has been successfully undertaken to perform sensor registration 
		in the context of distributed detection and tracking (DDT) of a single-target on a sensor network
		wherein a consensus Bernoulli filter \cite{guldogan2014consensus} is run at each sensor node.
		The novel contributions of this work include:
		a) the extension of sensor registration to the case of both unknown drift and orientation parameters, 
		which is impossible to accomplish in the context of \cite{gao2019distributed} (DDT of a single-target) due to the multiplicity of solutions for the orientation parameters;
		b) sensor registration is performed simultaneously with DMT, which represents a more general task for real-world applications.
		In this regard,
		if only the drift parameters are of interest and at most one target is present,
		the work of \cite{gao2019distributed} can be seen as a special case of this one.}


	\section{Problem Formulation and Background}
	\subsection{Problem Formulation}
	
	The aim of the paper is to jointly perform sensor registration and DMT
	using a time-synchronized sensor network. 
	Each node in the sensor network can get measurements of kinematic variables (e.g. angles, distances, Doppler shifts, etc.) relative to targets moving in the surrounding environment  
	and can process local data as well as exchange data with neighbors. 
	The network of interest has the following features: 
	it has no central fusion node; 
	sensor nodes are unaware of the network topology, i.e. the number of nodes and their connections; 
	each node maintains its own local coordinate frame
	and has no knowledge about the locations as well as azimuths of its neighbors with respect to its local coordinates.

	From a mathematical point of view, 
	the sensor network can be described in terms of a directed graph ${\cal G} = \left( {{\cal N},{\cal A}} \right)$, 
	where ${\cal N}$ is the set of sensor nodes and ${\cal A} \subseteq {\cal N} \times {\cal N}$ the set of connections such that $\left( {i,j} \right) \in {\cal A}$ if node $j$ can receive data from node $i$. 
	For each node $i \in {\cal N}$, 
	${{\cal N}^i}$ will denote the set of its in-neighbor nodes (including itself). 
	The total number of nodes in the network will be denoted by $\left| {\cal N} \right|$, i.e. the cardinality of $\cal N$. 
	
	It is assumed in this paper that each node expresses positions and velocities with respect to a local Cartesian coordinate frame
	and that, without loss of generality, each node is located at the origin of its own frame.
	Let $\left( {{\xi ^{i,j}},{\eta ^{i,j}}} \right)$ denote the position of node $j$ in the local coordinates of node $i$ where $j \in {{\cal N}^i}$,
	and define the drift parameter vector from node $j$ to $i$ as ${\vartheta ^{i,j}} = {\left[ {{\xi ^{i,j}}\;{\eta ^{i,j}}} \right]^\top }$.
	Similarly, ${\gamma ^{i,j}}$ is used to denote the orientation parameter from node $j$ to node $i$. 
	
	The single-target state expressed in the coordinates of node $i$ is denoted as
	${x^i} = {\left[ {{\xi ^i}\;{{\dot \xi }^i}\;{\eta ^i}\;{{\dot \eta }^i}} \right]^\top }$,
	where $\left( {\xi^i,\eta^i} \right)$ and $\left( {\dot \xi^i,\dot \eta^i} \right)$ are the target position and, respectively, velocity in Cartesian coordinates.
	It is easy to check that the target states $x^i$ and $x^j$ are related by
	\begin{equation}  \label{eq:TPT}
	{x^i} = {M^{i,j}}{x^j} + T{\vartheta ^{i,j}}
	\end{equation}
	where $T$ is the {\it transition matrix} defined as
	\begin{equation}
	T = {\left[ {\begin{array}{*{20}{c}}
			1&0&0&0\\
			0&0&1&0
			\end{array}} \right]^\top } \, ,
	\end{equation}
	$M \left( \gamma \right)$ is the {\it rotation matrix} defined as 
	
	\begin{align}
	M\left( \gamma  \right) = \left[ {\begin{array}{*{20}{c}}
		{\cos \left( \gamma  \right)}&0&{ - \sin \left( \gamma  \right)}&0\\
		0&{\cos \left( \gamma  \right)}&0&{ - \sin \left( \gamma  \right)}\\
		{\sin \left( \gamma  \right)}&0&{\cos \left( \gamma  \right)}&0\\
		0&{\sin \left( \gamma  \right)}&0&{\cos \left( \gamma  \right)}
		\end{array}} \right] \, ,
	\end{align}
	and, for convenience, the shorthand notation ${M^{i,j}} = M\left( {{\gamma ^{i,j}}} \right)$ is adopted.
	It is straightforward to check that the drift and orientation parameters satisfy the following properties
	\begin{align}
	{\vartheta ^{i,i}} &= 0, {\gamma ^{i,i}} = 0, {\gamma ^{i,j}} = - {\gamma ^{j,i}}, M^{i,i} = I_4  \\
	{\vartheta ^{i,j}} &= - T^\top{M^{i,j}}T{\vartheta ^{j,i}}   \\
	{M^{ - 1}}\left( \gamma  \right) &= {M^\top}\left( \gamma  \right) = M\left( { - \gamma } \right)  \\
	\det \left( {M\left( \gamma  \right)} \right) & \equiv \det \left( {{M^\top}\left( \gamma  \right)} \right) \equiv 1, \; \mbox{for  any}\; \gamma \, .
	\end{align}

\begin{figure}[t]
		\centering {
			\begin{tabular}{ccc}
				\includegraphics[width=0.40\textwidth]{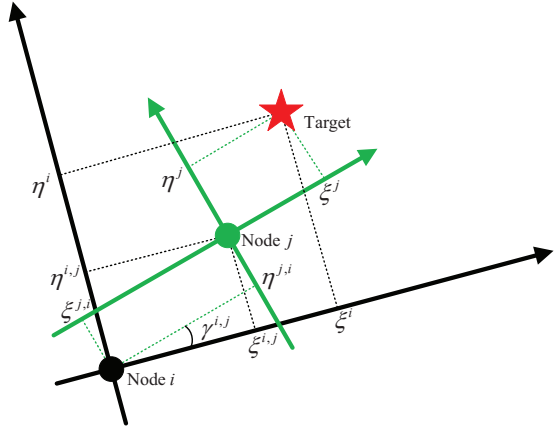}\\
			\end{tabular}
		}
		\caption{Drift and orientation between two sensor nodes.}
		\vspace{-0.5\baselineskip}
		\label{Fig:ExDOTN}
	\end{figure}

	In order to keep dimension consistence between the drift parameter and the target state vector, we define $\theta ^{i,j} = T \vartheta ^{i,j}$,
	then the drift parameter $\vartheta ^{i,j}$ can be easily recovered by computing $\vartheta ^{i,j} = T^\top\theta ^{i,j}$.
	The meaning of the drift and orientation (registration) parameters as well as the coordinate transformation (\ref{eq:TPT}) between two sensor nodes $i$ and $j$ are illustrated in
	Fig. \ref{Fig:ExDOTN}.
	
	The multitarget in the surveillance area at time $t$ is modeled as an RFS ${\cal X}_t$,
	which consists of $\left| {{{\cal X}}}_t \right|$ targets.
	Let us denote by ${\cal X}_t^i$ the multitarget RFS 
	expressed in the local coordinate frame of node $i$.
	The evolution of the target set ${{\cal X}_t}$ is supposed to be governed by the multitarget dynamics
	\begin{equation}
	{{\cal X}_{t + 1}} = {\Phi _{t + 1}}\left( {{{\cal X}_t}} \right)\bigcup {{{\cal B}_{t + 1}}}  \label{eq:EoMTD}
	\end{equation}
	where ${\cal B}_{t+1}$ is the RFS of new-born targets at time $t+1$ and
	\begin{align}
	{\Phi _{t + 1}}\left( {\cal X} \right) &= \bigcup\limits_{x \in {\cal X}} {{\phi _t}\left( x \right)}   \\
	{\phi _t}\left( x \right) &= \left\{ \begin{array}{l}
	\left\{ {{x_ + }} \right\},\;\; \rm {with\;survival\;probability} \; \it{P_{s,t}}\\
	\emptyset ,\;\;\;\;\;\;\;\;\;\rm{otherwise}
	\end{array} \right.   \label{eq:EoSTD}
	\end{align}
	with ${x_ + } $ distributed according to the single-target Markov transition PDF $\varphi_{t+1|t} ({x_ + } | x)$. 
	Notice that according to (\ref{eq:EoMTD})-(\ref{eq:EoSTD}) 
	each target in the set ${\cal X}_{t+1}$ is either a new-born target from the set ${\cal B}_{t+1}$ 
	or a target survived from ${\cal X}_t$, with probability $P_{s,t}$, 
	and whose state vector has evolved according to the single-target dynamics expressed by the transition PDF $\varphi_{t+1|t} (\cdot)$.
	In a similar way, observations are assumed to be generated, 
	at each node $i \in {\cal N}$, according to the measurement model
	\begin{equation}
	{\cal Y}_t^i = \Psi _t^i\left( {{\cal X}_t^i} \right)\bigcup {{\cal C}_t^i}   \label{eq:MTM}
	\end{equation}
	where ${\cal C}_t^i$ is the \emph{clutter} RFS (i.e. the set of measurements not due to targets) at time $t$ and node $i$, and
	\begin{align}
	\Psi _t^i\left( {{\cal X}^i} \right) &= \bigcup\limits_{x^i \in {\cal X}^i} {\psi _t^i\left( x \right)}   \\
	\psi _t^i\left( x^i \right) &= \left\{ \begin{array}{l}
	\left\{ y^i \right\}, \;\;\rm{with\;detection\;probability}\;\it{P_{d,t}^i}\\
	\emptyset ,\;\;\;\;\;\;\; \rm{otherwise}
	\end{array} \right.   
	\label{eq:STM}
	\end{align}
	with $y^i$ distributed according to the single-sensor likelihood $\mathcal L_t^i (y^i | x^i)$ at node $i$.
	Notice that according to (\ref{eq:MTM})-(\ref{eq:STM}) 
	each measurement in the set ${\cal Y}_t^i$ is either a false one from the clutter set ${\cal C}_t^i$ or is related to a target in ${\cal X}^i$, 
	with probability $P_{d,t}^i$, according to the single-sensor likelihood.
	
	The aim of this paper is, therefore, to estimate, at each node $i \in \mathcal{N}$,
	the drift and orientation parameters $\vartheta^{i,j}$ and $\gamma^{i,j}$, only for $j \in {\cal N}^i \backslash \left\{ i \right\}$
	(recall that each sensor can only communicate with neighbors)
	as well as the target set ${\cal X}_t^i$ 
	by collecting measurements and exchanging data with neighbors at each sampling interval.
	For convenience, let us also define ${\Theta ^i} = col\left( {{\theta ^{i,j}},j \in {{\cal N}^i}\backslash \left\{ i \right\}} \right)$,
	${{\cal M}^i} = { block-diag}\left( {M^{i,j},j \in {{\cal N}^i}\backslash \left\{ i \right\}} \right)$,
	$ {\Gamma ^i} = col\left( {{\gamma ^{i,j}},j \in {{\cal N}^i}\backslash \left\{ i \right\}} \right) $
	and ${{\cal T}^i} = block - diag\underbrace {\left( {T, \ldots ,T} \right)}_{\left| {{{\cal N}^i}} \right| - 1\;times}$.
	
	\subsection{Single-Sensor CPHD Filtering}
	
	From a probabilistic viewpoint, 
	an RFS $\cal X$ is completely characterized by its \emph{multitarget density} $f\left( \cal X \right)$.
	It is worth pointing out that the multiobject density, 
	while completely characterizing an RFS, 
	involves a combinatorial complexity; 
	hence simpler, though incomplete, 
	characterizations are usually adopted in order to keep the multitarget tracking problem computationally tractable. 
	In this paper, it is supposed that the multitarget RFS ${\cal X}$ is modelled by an \emph{i.i.d. cluster} point process with multitarget density of the form
	\begin{equation}
	f\left( {\cal X} \right) = \left| {\cal X} \right|! \,p\left( {\left| {\cal X} \right|} \right)\prod\limits_{x \in {\cal X}} {s\left( x \right)}   \label{eq:IIDCLU}
	\end{equation}
	where $p\left( n \right)$ is the \emph{probability mass function} (PMF) of the cardinality of ${\cal X}$ and $s\left( x \right)$ is the target spatial PDF.
	Clearly, an i.i.d. cluster point process is completely characterized by the pair $(p,s)$.
	
	
	The CPHD filter propagates in time the cardinality PMF ${p_{t}}\left( n \right)$  
	as well as the  target spatial PDF ${s_{t}}\left( x \right)$ of ${\cal X}_t$ 
	given ${{\cal Y}_{1:t }}$
	assuming that the clutter RFS, the predicted and filtered RFSs
	are i.i.d. cluster processes. 
	The resulting CPHD recursions (prediction and correction) can be found in \cite{mahler2007phd}.
	Note that, in principle, the PMF ${p_{t}}\left( n \right)$ is defined for a cardinality $n$ of the multitarget set going from $0$ to $\infty$;
	this is, of course, computationally infeasible.
	For implementation purposes,
	it is enough to assume a sufficiently large maximum number of targets $N_{max}$ in the scene.
	The spatial PDF ${s_{t}}\left( x \right)$  can be represented with \emph{particles} \cite{vo2005sequential} or as a \emph{Gaussian mixture} (GM) \cite{vo2007analytic}.
	In this paper, we adopt the GM representation of the CPHD filter, referred to as GM-CPHD filter, also used in \cite{battistelli2013consensus} for DMT.
	
	\subsection{Kullback-Leibler Paradigm for Multitarget Fusion}
	
	From an information-theoretic point of view, 
	fusion of multiple RFS densities through the network can be regarded as finding the RFS density 
	that minimizes the WAKLD among all the nodes of the sensor network \cite{battistelli2015distributed}. 
	To review the related concepts, 
	let us first introduce the notion of \emph{Kullback-Leibler Divergence} (KLD) from multitarget density $g\left( {\cal X} \right)$ to $f\left( {\cal X} \right)$ by
	
	\begin{equation}  \label{eq:KLD}
	{D_{KL}}\left( {\left. f \right\|g} \right) \buildrel \Delta \over = \int {f\left( {\cal X} \right)\log \frac{{f\left( {\cal X} \right)}}{{g\left( {\cal X} \right)}}\delta {\cal X}} 
	\end{equation}
	where the integral involved in (\ref{eq:KLD}) is the {\it set integral}
	defined in \cite{mahler2007statistical}.
	Then, the WKLA $\overline{f} \left( {\cal X} \right)$ of the RFS densities ${f^i}\left( {\cal X} \right)$ is defined as follows
	\begin{equation}  \label{eq:wKLA}
	\overline{f} \left( {\cal X} \right) \buildrel \Delta \over = \arg \mathop {\inf }\limits_f \underbrace{\sum\limits_i {{\omega ^i}{D_{KL}}\left( {\left. f \right\|{f^i}} \right)}}_{{\cal J}(f)}
	\end{equation}
	where the cost ${\cal J}(f)$ to be minimized with respect to $f(\cdot)$ is the WAKLD (Weighted Average Kullback-Leibler Divergence) and the weights ${\omega ^i}\geq 0$ must satisfy $\sum\nolimits_i {\omega ^i}  = 1$.
	In particular, if ${\omega ^i} = {1 \mathord{\left/
			{\vphantom {1 {\left| {\cal N} \right|}}} \right.
			\kern-\nulldelimiterspace} {\left| {\cal N} \right|}}$
	for $i = 1, \ldots ,\left| {\cal N} \right|$, (\ref{eq:wKLA}) provides the (unweighted) KLA which averages the node densities giving to all of them the same level of confidence. 
	An interesting interpretation of such a notion can be given recalling that, in Bayesian statistics, 
	the KLD (\ref{eq:KLD}) can be seen as the information gain achieved 
	when moving from a prior $g\left( {\cal X} \right)$ to a posterior $f\left( {\cal X} \right)$. 
	Thus, according to (\ref{eq:wKLA}), 
	the average density is the one that minimizes the weighted average of the information gains from the initial multitarget densities.
	
	\begin{theorem}\cite{battistelli2014kullback}  \label{the:wKLA}
		The WKLA defined in (\ref{eq:wKLA}) turns out to be given by
		\begin{equation}  \label{eq:fKLA}
		\overline{f} \left( {\cal X} \right) = \frac{{\prod\limits_i {{{\left[ {{f^i}\left( {\cal X} \right)} \right]}^{{\omega ^i}}}} }}{{\int {\prod\limits_i {{{\left[ {{f^i}\left( {\cal X} \right)} \right]}^{{\omega ^i}}}} \delta {\cal X}} }}
		\end{equation}
		and the  corresponding minimum WAKLD ${\cal J} \left( \overline{f} \right)$ is given by
		\begin{align}
		{\cal J} \left( \overline{f}  \right) &\buildrel \Delta \over = \sum\limits_i {{\omega ^i}{D_{KL}}\left( {\left. \overline{f} \right\|{f^i}} \right)}  \nonumber \\
		&=  - \log \left( {\int {\prod\limits_i {{{\left[ {{f^i}\left( {\cal X} \right)} \right]}^{{\omega ^i}}}} \delta {\cal X}} } \right)	\label{eq:COSKLA}
		\end{align}
	\end{theorem}
	
	Notice that (\ref{eq:fKLA}) corresponds to the normalized geometric mean of the densities $f^i$.
	Hence, Theorem \ref{the:wKLA} shows that the WKLA actually coincides with the GCI fusion rule,
	originally proposed by Mahler \cite{mahler2000optimal} as a generalization of
	Covariance Intersection to arbitrary densities.
	The minimal cost ${\cal J} \left( \overline{f} \right)$ in (\ref{eq:COSKLA}), which 
	is always nonnegative and vanishes only when all the densities are coincident, is known in the literature as
	{\em GCI divergence} \cite{suqi2018,suqi2019}. As discussed in \cite{suqi2018},
	the GCI divergence ${\cal J} \left( \overline{f} \right)$ makes it possible to quantify, in a principled way, the 
	degree of dissimilarity among a set of RFS densities within the context of GCI fusion.
	
	When all the densities to be fused are i.i.d. cluster densities, the WKLA can 
	be computed in closed form as follows.
	
	\begin{theorem} \cite{clark2010robust}
		Let all the $f^i$ be i.i.d. cluster densities characterized by the pairs $(p^i,s^i)$. Then, the  WKLA 
		$\overline{f} $ is again an i.i.d. cluster density characterized by the pair $(\overline{p},\overline{s})$ with
		\begin{align}
		\overline p\left( n \right) &= \frac{{\prod\limits_{i } {{{\left[ {{p^i}\left( n \right)} \right]}^{{\omega^i}}}{{\left( {\int {\prod\limits_{i} {{{\left[ {s^i\left( x \right)} \right]}^{{\omega ^i}}}} dx} } \right)}^n}} }}{{\sum\limits_{m = 0}^{{\infty}} {\prod\limits_{i} {{{\left[ {{p^i}\left( m \right)} \right]}^{{\omega^i}}}{{\left( {\int {\prod\limits_{i} {{{\left[ {s^i\left( x \right)} \right]}^{{\omega ^i}}}} dx} } \right)}^m}} } }}  \label{eq:OFCDCPHD}  \\
		\overline s^i\left( {{x}} \right){\rm{ }} &= \frac{{\prod\limits_{i}} {{{\left[ {s^i\left( x \right)} \right]}^{{\omega ^i}}} }}{{\int {\prod\limits_{i} {{{\left[ {s^i\left( x \right)} \right]}^{{\omega^i}}}} dx} }} \label{eq:OFSPCPHD} 
		\end{align}
	\end{theorem}
	
	In words, (\ref{eq:OFCDCPHD})-(\ref{eq:OFSPCPHD}) amount to state that the fusion of i.i.d. cluster processes provides
	an i.i.d. cluster process whose spatial PDF is the weighted geometric mean of the node spatial PDFs, 
	while the fused PMF is obtained
	by a more complicated expression (\ref{eq:OFCDCPHD}) also involving the node location PDFs besides the cardinality PMFs.

	\subsection{Distributed Multitarget Tracking}
	
	Let us preliminarily suppose that all the drift and orientation parameters ${\theta ^{i,j}},{\gamma ^{i,j}}\left( {i,j \in {\cal N}} \right)$ are known.
	Notice that this assumption is made here only for illustration purposes and will be relaxed later.
	
	When the drift and orientation parameters are known, 
	the ideas of Sections II.B and II.C can be combined so as to obtain an effective DMT algorithm.
	To see this, consider a generic time instant $t$ and suppose that,
	in each network node $i$,
	after the correction step of the GM-CPHD filter with local measurements,
	a RFS density $f_t^i\left( {{{\cal X}^i}} \right)$ is available representing the information at node $i$ on the i.i.d. cluster RFS ${\cal X}^i$ (expressed in the local coordinates of node $i$).
	Clearly, $f_t^i\left( {{{\cal X}^i}} \right)$ is characterized by the PMF $p_t^i\left( {\left| {{{\cal X}^i}} \right|} \right)$ and the spatial PDF $s_t^i\left( {{x^i}} \right)$.
	
	If all the densities $f_t^j\left( {{{\cal X}^j}} \right),j \in {\cal N}$, 
	were available in node $i$, fusion could be performed by:  1) expressing all the densities in the coordinates of node $i$ by means of the change of coordinates 
	(\ref{eq:TPT}) associated with ${\theta ^{i,j}},{\gamma ^{i,j}}\left( {i,j \in {\cal N}} \right)$; 2)
	computing the WKLA of such densities by means of the GCI fusion rule (\ref{eq:OFCDCPHD})-(\ref{eq:OFSPCPHD}). 
	
	Clearly, in a distributed setting, it is not possible to directly compute the fused density $\overline f_t$
	with (\ref{eq:OFCDCPHD})-(\ref{eq:OFSPCPHD})  since not all the densities $f_t^j\left( {{{\cal X}^j}} \right),j \in {\cal N}$,
	are available in node $i$. However, it turns out that the collective average $\overline f_t$ can be 
	approximated to any desired degree of accuracy by means of distributed computation
	(i.e. exchanging only information with the neighbors).
	This is made possible by the \emph{consensus method}
	which has emerged as a powerful tool for distributed computation
	over networks and has found widespread applications, e.g., in distributed parameter/state estimation.
	In essence, consensus aims at computing the collective average
	by iterating several times the computation of the regional average over the sub-network ${\cal N}^i$
	of in-neighbors of each node $i$.
	In fact, it can be shown that,
	under suitable conditions, 
	as the number $L$ of consensus iterations increases, 
	the density in each node converges to the collective average $\overline f_t$ \cite{battistelli2013consensus,battistelli2014kullback}.
	
	Then, in practice, at each time $t$ each node $i$ of the network,
	after local GM-CPHD filtering, 
	iterates for $L$ times data-exchange with the neighbors and fusion of the received densities with the local one.
	More specifically, consider a generic node $i$ at time $t$ and 
	suppose that $\ell$ consensus iterations have been carried out.
	Then, the density at node $i$ expressed in local coordinates is 
	an i.i.d. cluster density characterized by the pair $(p_{t,\ell }^i, s_{t,\ell }^i)$.
	
	Notice that each $s_{t,\ell }^j$ is expressed in the coordinates of node $j$, that is, it is a function of $x^j$. In order to compute the regional average over the sub-network ${\cal N}^i$, each node $i$ applies the changes of coordinates (\ref{eq:TPT}) 
	to the spatial PDFs of the neighbors so as to obtain the densities
	$ s_{t,\ell }^{j,i} (x^i ; \theta^{i,j} , \gamma^{i,j}) $
	for $j \in \mathcal N_i$
	{where, clearly,
		$s_{t,\ell }^{j,i}\left( {{x^i};{\theta ^{i,j}},{\gamma ^{i,j}}} \right) = s_{t,\ell }^i\left( {{M^{i,j}}{x^i} + T{\theta ^{i,j}}} \right)$
		and 
		$s_{t,\ell }^{i,i} = s_{t,\ell }^{i} $.}
	Then, the fused density at the next consensus step $f_{t,\ell  + 1}^i\left( {{{\cal X}^i}} \right)$ is an 
	i.i.d. cluster density characterized by the pair $(p_{t,\ell+1 }^i, s_{t,\ell+1 }^i)$ where
	\begin{align}
	p_{t,\ell  + 1}^i\left( n \right) 
	& = \frac{{\prod\limits_{j \in {{\cal N}^i}} {{{\left[ {p_{t,\ell }^j\left( n \right)} \right]}^{{\omega ^{i,j}}}}{{\left( {\int {\prod\limits_{j \in {{\cal N}^i}} {{{\left[ {s_{t,\ell }^{j,i} \left( x ; \theta^{i,j} , \gamma^{i,j} \right)} \right]}^{{\omega ^{i,j}}}}} dx} } \right)}^n}} }}{{\sum\limits_{m = 0}^{{\infty}} {\prod\limits_{j \in {{\cal N}^i}} {{{\left[ {p_{t,\ell }^j\left( n \right)} \right]}^{{\omega ^{i,j}}}}{{\left( {\int {\prod\limits_{j \in {{\cal N}^i}} {{{\left[ {s_{t,\ell }^{j,i}\left( x ; \theta^{i,j} , \gamma^{i,j} \right)} \right]}^{{\omega ^{i,j}}}}} dx} } \right)}^m}} } }} \label{eq:DCPMF}  \\
	s_{t,\ell  + 1}^i \left( x^i \right) &= \frac{{\prod\limits_{j \in \mathcal N^i}} {{{\left[ {s^{j,i}_{t,\ell}\left( x^i ;   \theta^{i,j} , \gamma^{i,j}\right)} \right]}^{{\omega ^{i,j}}}} }}{\int {\prod\limits_{j \in \mathcal N^i}} {{{\left[ {s^{j,i}_{t,\ell}\left( x ;   \theta^{i,j} , \gamma^{i,j}\right)} \right]}^{{\omega ^{i,j}}}} dx} } 
	\label{eq:DCPHDC}
	\end{align}

	The CCPHD filter is summarized in Table \ref{tab:GCICPHD}. For a practical implementation of such an algorithm based on a GM approximation of the
	spatial PDFs $s_{t,\ell }^i$, the interested reader is referred to \cite{battistelli2013consensus}.
	
	
	\setcounter{magicrownumbers}{0}
	\begin{table}
		\caption{CCPHD Filter (node $i$, time $t$)}   \label{tab:GCICPHD}
		\begin{center}
			\begin{tabular}{l p{7cm}}
				\hline \hline
				\textbf{Input:} & $f_{t - 1}^i\left( {{{\cal X}^i}} \right)$ and ${\left( {{\theta ^{i,j}},{\gamma ^{i,j}}} \right),j \in {{\cal N}^i}}$   \\
				\hline \hline
				\rownumber$\;$ & Local prediction and correction of the CPHD filter to get the pair
				$(p^i_{t},s^i_{t})$ \\ 
				\rownumber$\;$ & Set $p_{t,0}^i = p_{t}^i $ and $s_{t,0}^i = s_{t}^i $ \\ 
				\rownumber$\;$ & For $\ell  = 0, \ldots ,L-1$, do \\
				\rownumber$\;$ & $\;\;\;\;$ Exchange information with the neighbors $j \in {\cal N}^i \backslash \left\{ i \right\}$ \\
				& $\;\;\;\;$ to get the pairs $(p^j_{t,\ell},s^j_{t,\ell})$ \\
				\rownumber$\;$ & $\;\;\;\;$ Change of coordinates to get the spatial PDFs $s^{j,i}_{t,\ell}$  \\                       
				\rownumber$\;$ & $\;\;\;\;$ GCI Fusion using (\ref{eq:DCPMF})-(\ref{eq:DCPHDC})  \\
				\rownumber$\;$ & End \\
				\rownumber$\;$ & Set $p_{t}^i = p_{t,L}^i $ and $s_{t}^i = s_{t,L}^i $ \\
				\rownumber$\;$ & Multitarget State Estimation, i.e. first estimate the number of targets according to the PMF $p_t^i$, and then extract the corresponding number of peaks from the spatial PDF $s_t^i\left( x \right)$ \\
				\hline \hline
			\end{tabular}
		\end{center}
	\end{table}

	\section{Distributed Sensor Self-localization}
	
	\subsection{The GCI divergence for i.i.d. cluster RFS densities}
	
	The previous section has introduced the CCPHD filter,
	which assumes that the drift and orientation parameters between node $i$ and $j$, 
	$\left( {i,j} \right) \in {\cal A}$,
	are known a priori. 
	However, in practice, registration parameters may not be known, or at least be known with insufficient accuracy. 
	Then ${\theta ^{i,j}}$ and $\gamma ^{i,j}, \left( {i,j} \right) \in {\cal A}$, need to be estimated together or as a premise to the target RFS $\cal X$
	without additional localization hardware in the sensor nodes and in the surrounding environment.
	
	The sensor registration approach proposed  in this section is suboptimal but has the twofold advantage of being scalable 
	and not requiring any global information on the network topology. 
	As a further benefit, in order to keep the communication load as low as possible, 
	the proposed approach will not require any additional data exchange with respect to the CCPHD filter of Table \ref{tab:GCICPHD}.
	
	The idea is to exploit the information-theoretic interpretation of the consensus step (\ref{eq:DCPMF})-(\ref{eq:DCPHDC})
	in order to define a suitable cost function which can be used for estimation of the registration parameters. 
	As discussed in the previous section, each consensus step in node $i$ amounts to computing a regional average, according to the WKLA paradigm, 
	over the subnetwork $\mathcal N^i$ of in-neighbors of node $i$. Then, as explained in Section III-C, a natural way of measuring the discrepancy
	among the multitarget densities to be fused is the minimal cost after fusion, that is the GCI divergence
	\begin{align}
	{\cal J}_{t,\ell }^i\left( {{\Theta ^i},{\Gamma ^i}} \right)   \buildrel \Delta \over =   - \log \left\{ {\int {\prod\limits_{j \in {{\cal N}^i}} {{{\left[ {f_{t,\ell  - 1}^{j,i}\left( {{\cal X}^i;{\theta ^{i,j}},{\gamma ^{i,j}}} \right)} \right]}^{{\omega ^{i,j}}}}} \delta {\cal X}} } \right\}   \nonumber 
	\end{align}
	where  ${f_{t,\ell  - 1}^{j,i}\left( {{\cal X}^i;{\theta ^{i,j}},{\gamma ^{i,j}}} \right)}$ represents the multitarget density of node $j$ expressed in the coordinates of node $i$.
	Accordingly, such a quantity represents the \emph{instantaneous cost} (IC) of node $i$ at time $t$  and consensus step $\ell$ to be minimized in order to estimate
	drift ${\theta ^{i,j}}$ and orientation $\gamma ^{i,j}$ parameters for any $j \in \mathcal N^i \setminus \{i\}$. 
	The rationale for such a choice is that, 
	when all the local filters perform well, 
	then all the local densities should provide a reasonably accurate estimate of the target set in local coordinates. 
	In this case, the discrepancy between the multitarget densities
	in two neighboring nodes $i$ and $j$ is mainly due to the different coordinates. 
	Hence, it is reasonable to take as estimate of the drift ${\theta ^{i,j}}$ and orientation $\gamma ^{i,j}$ parameters
	the values which minimize such a discrepancy. 
	
	Since in the considered setting all the densities to be fused are i.i.d. cluster densities, the IC ${\cal J}_{t,\ell }^i\left( {{\Theta ^i},{\Gamma ^i}} \right) $
	can be further specified as follows (see Appendix A for the proof).
	
	\begin{proposition} Let all the local multitarget densities $f_{t,\ell  - 1}^j$, $j \in \mathcal N^i$, be i.i.d. cluster densities characterized by the pair $(p_{t,\ell-1}^j, s_{t,\ell-1}^j)$. 
		Then, the IC can be computed as follows
		\begin{equation}
		{\cal J}_{t,\ell }^i\left( {{\Theta ^i},{\Gamma ^i}} \right)   = - \log \left\{ {\sum\limits_{n = 0}^{{\infty}} {c_{t,\ell}^{i,n}{{\left[ {{\cal W}_{t,\ell }^i\left( {{\Theta ^i},{\Gamma ^i}} \right)} \right]}^n}} } \right\}  \label{eq:CoCDGMCPHD}
		\end{equation}
		where
		\begin{align}
		c_{t,\ell}^{i,n} &= \prod\limits_{j \in {{\cal N}^i}} {{{\left[ {p_{t,\ell  - 1}^j\left( n \right)} \right]}^{{\omega ^{i,j}}}}}  \label{eq:CIIF}  \\
		{\cal W}_{t,\ell }^i\left( {{\Theta ^i},{\Gamma ^i}} \right) &= \int {\prod\limits_{j \in {{\cal N}^i}} {{{\left[ {s_{t,\ell  - 1}^{j,i}\left( {x;{\theta ^{i,j}},{\gamma ^{i,j}}} \right)} \right]}^{{\omega ^{i,j}}}}dx} }   \label{eq:IRF}
		\end{align}
		\label{prop1}
	\end{proposition}
	
	It can be seen that $c_{t,\ell}^{i,n}$ is a constant independent of both the drift and orientation parameters.
	Conversely, ${\cal W}_{t,\ell }^i\left( {{\Theta ^i},{\Gamma ^i}} \right)$, referred to hereafter as \emph{instantaneous reward factor} (IRF) of node $i$ at time $t$ and consensus step $\ell$,
	is the only part of the IC  related to the drift and orientation parameters.
	Please notice that minimizing the IC ${\cal J}_{t,\ell }^i\left( {{\Theta ^i},{\Gamma ^i}} \right)$  is the same as maximizing the IRF ${\cal W}_{t,\ell }^i\left( {{\Theta ^i},{\Gamma ^i}} \right)$, with respect to the registration parameters $\Theta^i$ and $\Gamma^i$. 
	
	
	\subsection{Gaussian Mixture implementation}
	
	In this section, we discuss how the IRF can be computed when a GM implementation of the CCPHD filter is adopted. Of course, this amounts to assuming that 
	all the spatial densities $s_{t,\ell-1}^i$ are represented as GMs, i.e.,
	\begin{align}
	s_{t,\ell-1}^i\left( {{x^i}} \right) = \sum\limits_{k = 1}^{N_{t,\ell-1}^i} {\alpha_{t,\ell-1}^{i,k} ~{\cal G}\left( {{x^i};\mu _{t,\ell-1}^{i,k},P_{t,\ell-1}^{i,k}} \right)}
	\label{eq:GM}
	\end{align}
	where $N_{t,\ell-1}^i$ is the number of Gaussian components, the weights $\alpha_{t,\ell-1}^{i,k}$  of the mixture are positive and such that
	$\sum_{k = 1}^{N_{t,\ell-1}^i} \alpha_{t,\ell-1}^{i,k} = 1 $, and ${\cal G} (x;\mu,P ) $ denotes a Gaussian PDF with mean $\mu$ and covariance matrix $P$.
	
	Preliminary operations for the computation of the IRF (\ref{eq:IRF}) in node $i$ are: the exponentiation by $\omega^{i,j}$ of each $s_{t,\ell-1}^j$, $j \in \mathcal N^i$;
	and the application of the change of coordinates (\ref{eq:TPT}) for any $j \in \mathcal N^i \setminus \{i\}$. In this paper, following \cite{battistelli2013consensus,gunay2016chernoff},
	the power of each $s_{t,\ell-1}^j$ is supposed to be approximated as a GM of the form
	\begin{align}
	{\left[s_{t,\ell-1 }^j \left( x^j \right) \right]^{{\omega ^{i,j}}}} \cong \sum\limits_{k = 1}^{N_{t,\ell-1}^j} { {\widehat \alpha} _{t,\ell-1}^{j,i,k} \, {\cal G}\left( {x^j;
			\mu _{t,\ell-1 }^{j,k},
			P _{t,\ell-1 }^{j,k}}  / \omega^{i,j} \right)}    \label{eq:AOPGM}
	\end{align}
	There are several methods to determine the weight of each Gaussian component, 
	such as the computationally-cheap one adopted in \cite{battistelli2013consensus} 
	or the newly proposed method of \cite{gunay2016chernoff}.
	However, the choice of a particular approximation method is immaterial for the subsequent developments as long as an approximation like (\ref{eq:AOPGM}) holds.
	Concerning the change of variables, it is immediate to check that each Gaussian in (\ref{eq:AOPGM}) can be rewritten as
	\begin{equation}\label{eq:CoC}
	{\cal G}\left( {x^j; \mu _{t,\ell-1 }^{j,k}, P _{t,\ell-1 }^{j,k}}  / \omega^{i,j} \right)
	= {\cal G}\left( {x^i; \widehat \mu _{t,\ell-1 }^{j,i,k}, \widehat P _{t,\ell-1 }^{j,i,k}} \right)
	\end{equation}
	where
	\begin{align}
	{\widehat \mu} _{t,\ell-1}^{j,i,k} &= {M^{i,j}}\mu _{t,\ell -1}^{j,k} + {\theta ^{i,j}}   \\
	{\widehat P} _{t,\ell-1}^{j,i,k} &= {{{M^{i,j}} P_{t,\ell -1}^{j,k}{{\left( {{M^{i,j}}} \right)}^\prime }}}/{{{\omega ^{i,j}}}}
	\end{align}
	
	Hence, computation of the IRF (\ref{eq:IRF}) in node $i$ involves the product of $| \mathcal N^i |$ GMs. With this respect, 
	we observe that such a product is again a GM having a total of $ \prod\nolimits_{j \in {{\cal N}^i}} {N_{t,\ell  - 1}^i} $ Gaussian components. 
	Hereafter, for the sake of compactness, we will denote each of such components by means of a vector index, say $k$, taking value in the set 
	\[
	\mathcal K_{t,\ell}^i = \mathop \times \limits_{j \in {\mathcal N^i}} \left\{ 1, \ldots ,N_{t,\ell-1}^j \right\} 
	\]
	where, here, $\times$ denotes Cartesian product. Accordingly, each element $k(j)$, $j \in \mathcal N^i$, of the vector index $k \in \mathcal K_{t,\ell}^i $
	expresses which of the Gaussian components of node $j$ is used to form the $k$-th component of the product.
	
	Taking into account the above considerations and notation, the following result can now be stated. 
	
	
	\begin{theorem}  \label{prop:CALI} Let all the local spatial densities be represented by GMs as in (\ref{eq:GM}) and let the approximation (\ref{eq:AOPGM}) be adopted.
		Then the IRF turns out to be a GM given by
		\begin{align}
		{\cal W}_{t,\ell }^i\left( {{\Theta ^i},{\Gamma ^i}} \right)  = \sum\limits_{k \in \mathcal K_{t,\ell}^i } {\beta _{t,\ell }^{i,k}\left( {{\Gamma ^i}} \right){\cal G}\left( {{\Theta ^i};\phi _{t,\ell }^{i,k}\left( {{\Gamma ^i}} \right),\Upsilon _{t,\ell }^{i, k}\left( {{\Gamma ^i}} \right)} \right)}
		\label{eq:SubCP}
		\end{align}
		where	
		\begin{eqnarray}
		\beta _{t,\ell }^{i, k}\left( {{\Gamma ^i}} \right) &= & \det {\left[ {2\pi \overline P_{t,\ell }^{i,k}\left( {{\Gamma ^i}} \right)} \right]^{\frac{1}{2}}}\det {\left[ {2\pi \Upsilon _{t,\ell }^{i, k}\left( {{\Gamma ^i}} \right)} \right]^{\frac{1}{2}}}    \prod\limits_{j \in {{\cal N}^i}} {\frac{{  {\widehat \alpha } _{t,\ell-1}^{j,i, k\left( j \right)}}}{{\det {{\left( {2\pi {{P_{t,\ell  - 1}^{j, k\left( j \right)}} \mathord{\left/
										{\vphantom {{P_{t,\ell  - 1}^{j, k\left( j \right)}} {{\omega ^{i,j}}}}} \right.
										\kern-\nulldelimiterspace} {{\omega ^{i,j}}}}} \right)}^{\frac{1}{2}}}}}}  \label{eq:Bphi} \\
		\phi _{t,\ell }^{i, k}\left( {{\Gamma ^i}} \right) &=& {E^i} \mu _{t,\ell  - 1}^{i,k\left( i \right)} - {{\cal M}^i}{\bf{u}}_{t,\ell  - 1}^{i, k}  \label{eq:IRFME} \\
		\Upsilon _{t,\ell }^{i, k}\left( {{\Gamma ^i}} \right) &=& { \Psi _{t,\ell }^{i, k} } + {E^i}{{P_{t,\ell  - 1}^{i,k\left( i \right)}}}{\left( {{E^i}} \right)^\top / \omega ^{i,i}}
		\end{eqnarray}
		and	
		\begin{eqnarray}
		\hspace{-.4cm} \Psi _{t,\ell }^{i, k} &=& \hspace{-.2cm} block-diag\left[   \widehat P_{t,\ell  - 1}^{j,k  ( j )}  , \, j \in {{\cal N}^i} \backslash \left\{ i \right\} \right] \\
		\hspace{-.4cm} {E^i} &=& \hspace{-.2cm}  col\underbrace {\left( {{I_4}, \ldots ,{I_4}} \right)}_{\left| {{{\cal N}^i}} \right| - 1\;times} \\
		{\bf{u}}_{t,\ell  - 1}^{i,k} &=& col\left( {\mu _{t,\ell  - 1}^{j,k\left( j \right)},j \in {{\cal N}^i}\backslash \left\{ i \right\}} \right) \\
		\overline P_{t,\ell }^{i, k}\left( {{\Gamma ^i}} \right) &=& {\left[ {{{\left( {{E^i}} \right)}^\top } \left ( \Psi _{t,\ell }^{i, k} \right )^{-1} {E^i} + {{\left( {{{P_{t,\ell  - 1}^{i, k\left( i \right)}}}/{{{\omega ^{i,i}}}}} \right)}^{ - 1}}} \right]^{ - 1}}  \label{eq:FCFC}
		\end{eqnarray}

	\end{theorem}
	
	The proof of Theorem \ref{prop:CALI} is given in Appendix A.
	Notice that the IC defined in (\ref{eq:CoCDGMCPHD}) is minimized if and only if the sensor nodes are correctly registered,
	in which case the IRF ${\cal W}_{t,\ell }^i\left( {{\Theta ^i},{\Gamma ^i}} \right)$ is maximized.
	In order to take a further look at how the IRF can be used for sensor registration purposes,
	the following remarks are in order.
	
	\begin{remark}  \label{RE:PI}
		Notice that (\ref{eq:SubCP}) consists of $ | \mathcal K_{t,\ell}^i | = \prod\nolimits_{j \in {{\cal N}^i}} {N_{t,\ell  - 1}^i} $ Gaussian components,
		where each component $k \in  \mathcal K_{t,\ell}^i $ represents a possible association
		among Gaussian components $k(j)$, $ j \in \mathcal N^i$, (which represent potential targets) of spatial PDFs from neighboring nodes.
		Generally speaking, correct associations among Gaussian components should approximately have the same drift and orientation parameters
		(i.e. the means ${\phi _{t,\ell }^{i,k}\left( {{\Gamma ^i}} \right)}$ of Gaussian components which represent correct target associations should be almost the same).
		By recognizing this, a convenient optimization approach to maximize ${\cal W}_{t,\ell }^i\left( {{\Theta ^i},{\Gamma ^i}} \right)$ can be adopted,
		as it will be discussed in the next sections.
	\end{remark}
	
	\begin{remark} \label{RE:NAP}
		When the orientation parameters are known a priori (e.g., the static sensor nodes are deployed to point towards the same direction),
		the weight, mean and covariance of each Gaussian component in (\ref{eq:SubCP}) turn out to
		be constant, i.e. independent of the orientation parameters $\Gamma ^i$,
		which means that the maximum of  the IRF can be found by directly operating on GMs \cite{ari2012maximum,karlis2003choosing}.
	\end{remark}
	
	\begin{remark} \label{RE:SMC}
		{
			It should be noticed that
			the IRF ${\cal W}_{t,\ell }^i\left( {{\Theta ^i},{\Gamma ^i}} \right)$ cannot be directly computed
			when the spatial PDFs $s_{t,\ell }^i\left(  \cdot  \right),i \in {\cal N}$, are represented by particles,
			i.e. $s_{t,\ell }^i\left( {{x^i}} \right) = \sum\limits_{k = 1}^{N_{t,\ell }^i} {\alpha _{t,\ell }^{i,k}{\delta _{\mu _{t,\ell }^{i,k}}}\left( {{x^i}} \right)}$
			where
			\[
			{\delta _\mu }\left( x \right) = \left\{ \begin{array}{l}
			1,\;x = \mu \\
			0,\;{\rm otherwise}
			\end{array} \right.
			\]
			denotes the \emph{Dirac delta}.
			This is
			due to the fact that each sensor node in the network
			would locally store and propagate its own set of particles,
			so that (\ref{eq:IRF}) would always be equal to zero.
			A viable approach is to use a GM to approximate the (delta-mixture) particle representation (e.g., ,
			by adopting the method of \cite{sheng2005distributed}), then fuse GMs and finally transform the fused GM back to particle 
			representation.
		}
	\end{remark}

	\subsection{Sensor registration with known orientation parameters}
	
	In this section, it is assumed that the orientation parameters between any pair of neighbor nodes are known a priori
	and that, without loss of generality, ${\gamma ^{i,j}} = 0$ for any $\left( {i,j} \right) \in {\cal A}$. 
	In this case, ${M^{i,j}} = {I_4}$ so that (\ref{eq:SubCP}) can be rewritten as
	\begin{align}
	{\cal W}_{t,\ell }^i\left( {{\Theta ^i}} \right) = \sum\limits_{ k \in \mathcal K_{t,\ell}^i} {\beta _{t,\ell }^{i, k} ~{\cal G}\left( {{\Theta ^i};\phi _{t,\ell }^{i, k},\Upsilon _{t,\ell }^{i,k}} \right)}   \label{eq:IRFKD}
	\end{align}
	where the weight ${\beta _{t,\ell }^{i, k}}$, mean ${\phi _{t,\ell }^{i, k}}$ and covariance ${\Upsilon _{t,\ell }^{i,k}}$ of each Gaussian component can be computed through
	(\ref{eq:Bphi})-(\ref{eq:FCFC}) by setting ${{\cal M}^i} = {I_{4 (\left| {{{\cal N}^i}} \right | -1)}}$.
	Hence, the instantaneous cost (\ref{eq:CoCDGMCPHD}) is independent of $\Gamma ^i$.
	
	Since minimization of the IC ${\cal J}_{t,\ell }^i\left( {{\Theta ^i}} \right)$, or equivalently maximization of the IRF
	$\mathcal W_{t,\ell}^i (\Theta^i)$,
	would make the estimate too sensitive to transient errors in the local densities,
	in order to obtain a more reliable estimate of the drift parameters 
	we follow a strategy usually adopted in recursive parameter estimation and 
	consider a \emph{total cost} (TC) up to the current time instant/consensus step
	\begin{align}
	{\cal J}_{1:t,\ell }^i\left( {{\Theta ^i}} \right) = \sum\limits_{\tau  = 1}^{t - 1} {\sum\limits_{l = 1}^L {{\cal J}_{\tau ,l}^i\left( {{\Theta ^i}} \right)} }  + \sum\limits_{l = 1}^\ell  {{\cal J}_{t,l}^i\left( {{\Theta ^i}} \right)}   \label{eq:TC}
	\end{align}
	In the sequel, an algorithm is provided for addressing minimization of the TC ${\cal J}_{1:t,\ell }^i\left( {{\Theta ^i}} \right) $ when
	a GM implementation of the CCPHD filter is adopted.
	
	To this end, observe first that, in principle, computation of the IC ${\cal J}_{t,\ell }^i\left( {{\Theta ^i}} \right)$ as in (\ref{eq:CoCDGMCPHD}) would involve the summation of an infinite number of GMs (all the powers of the IRF $\mathcal W_{t,\ell}^i (\Theta^i)$ for $n$ going to infinity). However, in practice, for implementation purposes in the GM-CCPHD filter
	the PMFs $p^j_{t,\ell-1} (n)$, and hence the scalars $c_{t,\ell}^{i,n}$, are all equal to zero for $n$ greater than  $N_{max}$, the assumed maximum number of targets in the scene \cite{vo2007analytic}.
	Hence, at most $N_{max}$ GMs have to be taken into account in the summation so that the IC can be written as 
	\begin{align}
	{\cal J}_{t,\ell}^i\left( {{\Theta ^i}} \right) =  - \log \left[ {c_{t,\ell}^{i,0} + \widetilde {\cal W}_{t,\ell }^i\left( {{\Theta ^i}} \right)} \right]   \label{eq:ICN}
	\end{align}
	where ${c_{t,l}^{i,0}}$ is given in (\ref{eq:CIIF}), 
	and 
	\begin{equation}\label{eq:W-tilde}
	\widetilde {\cal W}_{t,\ell }^i \left( {{\Theta ^i}} \right) =  \textstyle{\sum_{n=1}^{N_{max}} c_{t,\ell}^{i,n} \left [ \mathcal W _{t,\ell}^i (\Theta^i) \right ]^n}
	\end{equation}
	is a GM. Then, it is an easy matter to check that also the TC can be written as 
	\begin{align}
	{\cal J}_{1:t,\ell }^i\left( {{\Theta ^i}} \right) =  - \log \left[ {{\cal C}_{1:t,\ell}^i + \widetilde {\cal W}_{1:t,\ell }^i\left( {{\Theta ^i}} \right)} \right]
	\end{align}
	where ${\cal C}_{1:t,\ell}^i = \prod\nolimits_{l = 1}^\ell  {c_{t,l}^{i,0}} \left( {\prod\nolimits_{\tau  = 1}^{t - 1} {\prod\nolimits_{l = 1}^L {c_{\tau ,l}^{i,0}} } } \right)$
	and ${\widetilde {\cal W}_{1:t,\ell }^i\left( {{\Theta ^i}} \right)}$ is again a GM. Further, ${\cal C}_{1:t,\ell}^i$ and $\widetilde {\cal W}_{1:t,\ell }^i\left( {{\Theta ^i}} \right)$ can be
	recursively computed as follows
	\begin{align}
	{\cal C}_{1:t,\ell}^i &= {\cal C}_{1:t,\ell-1}^i c_{t,\ell}^{i,0}   \label{eq:rec1} \vspace{2mm} \\
	\widetilde {\cal W}_{1:t,\ell}^i\left( {{\Theta ^i}} \right) &= c_{t,\ell}^{i,0} \, \widetilde {\cal W}_{1:t,\ell-1}^i\left( {{\Theta ^i}} \right) + {\cal C}_{1:t,\ell-1}^i \, \widetilde {\cal W}_{t,\ell }^i\left( {{\Theta ^i}} \right)  + \widetilde {\cal W}_{t,\ell }^i\left( {{\Theta ^i}} \right)\widetilde {\cal W}_{1:t,\ell-1}^i\left( {{\Theta ^i}} \right)  \label{eq:rec2}
	\end{align}
	Clearly, in the computation of $\widetilde {\cal W}_{1:t,\ell}^i\left( {{\Theta ^i}} \right)$ \emph{merging and pruning} techniques can be adopted in order to
	keep the number of Gaussian components below a pre-specified threshold, thus ensuring bounded complexity as $t$ and $\ell$ increase.
	

	Hence, the estimated drift parameters at time $t$ and consensus step $\ell$ are obtained by solving the following optimization 
	problem\footnote{Recalling that, by construction, $\Theta ^i = col \left (T \vartheta^{i,j} , j \in \mathcal N^i \setminus \{i\} \right) $, when writing
		(\ref{opt}) we intend that the optimization is performed with respect to the parameters $\vartheta^{i,j} $ and $\gamma^{i,j}$ with $j \in \mathcal N^i \setminus \{i\})$.
	}
	\begin{align}
	\widehat \Theta _{1:t,\ell }^i = \mathop {\arg} \min\limits_{{\Theta ^i}} {\cal J}_{1:t,\ell }^i\left( {{\Theta ^i}} \right) =
	\mathop {\arg} \max \limits_{{\Theta ^i}} \widetilde {\cal W}_{1:t,\ell}^i\left( {{\Theta ^i}} \right) \label{opt}
	\end{align}
	which amounts to finding the global maximum of a GM.
	Candidate methods to solve (\ref{opt}) are, e.g., \emph{grid search}, which first searches for a initial point and is then followed by a \emph{gradient-based algorithm}, 
	and \emph{multiple-initialization} which is run in parallel from multiple initial points \cite{ari2012maximum,karlis2003choosing}.
	Note that, after several sampling times 
	when the estimation of the multitarget states at sensor nodes becomes stationary,
	one can directly employ the estimated drift parameters at time $t$, 
	consensus step $\ell-1$ (i.e. $\widehat \Theta _{1:t,\ell-1 }^i$) as  initial point.
	The proposed sensor registration method with known orientation parameters is summarized in Table \ref{tab:SRKO}.
	
	\setcounter{magicrownumbers}{0}
	\begin{table}
		\caption{Sensor registration with known orientation (node $i$, time $t$, consensus iteration $\ell$)}   \label{tab:SRKO}
		\begin{center}
			\begin{tabular}{l p{9.5cm}}
				\hline \hline
				\textbf{Input:}  &
				$\widetilde {\cal W}_{1:t,\ell-1}^i\left( {{\Theta ^i}} \right)$, ${\cal C}_{1:t,\ell-1}^i$, and			
				$f_{t,\ell  - 1}^j\left( {\cal X}^j \right),j \in {{\cal N}^i}$
				\\
				\hline \hline		
				\rownumber & Compute the IRF ${\cal W}_{t,\ell }^i\left( {{\Theta ^i}} \right)$  using (\ref{eq:SubCP})-(\ref{eq:FCFC}) \\		
				\rownumber & Compute  $\widetilde {\cal W}_{t,\ell }^i$ and ${c_{t,l}^{i,0}}$ using (\ref{eq:W-tilde}) and (\ref{eq:CIIF}) \\
				\rownumber & Compute  $\widetilde {\cal W}_{1:t,\ell}^i\left( {{\Theta ^i}} \right)$ and ${\cal C}_{1:t,\ell}^i$ using (\ref{eq:rec1})-(\ref{eq:rec2}) \\
				\rownumber & Perform pruning and merging on $\widetilde {\cal W}_{1:t,\ell }^i\left( {{\Theta ^i}} \right)$ \\
				\rownumber & Find $ {\widehat \Theta _{1:t,\ell} ^i}$ by maximizing $\widetilde {\cal W}_{1:t,\ell }^i\left( {{\Theta ^i}} \right)$ using the \textit{multiple initialization} strategy \\	
				\hline \hline
				\textbf{Output}: &
				${\widehat \Theta _{1:t,\ell} ^i}$, $\widetilde {\cal W}_{1:t,\ell}^i\left( {{\Theta ^i}} \right)$, ${\cal C}_{1:t,\ell}^i$\\
				\hline \hline
			\end{tabular}
		\end{center}
	\end{table}
	
	\subsection{Sensor registration with both unkown drift and orientation parameters}
	
	When both drift and orientation parameters are unknown,
	the TC defined in (\ref{eq:TC}) will actually become a function ${\cal J}_{1:t,\ell }^i\left( {{\Theta ^i},{\Gamma ^i}} \right)$ of both  $\Theta ^i$ and ${\Gamma ^i}$.
	In this case, direct optimization of the TC will become difficult
	since, due to the dependence on $\Gamma^i$ of the mean ${\phi _{t,\ell }^{i,k}\left( {{\Gamma ^i}} \right)}$ 
	and covariance ${\Upsilon _{t,\ell }^{i, k}\left( {{\Gamma ^i}} \right)}$ 
	of each Gaussian component of the IRF,
	\emph{merging and pruning} strategies cannot be directly implemented,
	thus implying an exponential increase in the number of Gaussian components of the TC.
	In order to overcome such a drawback, in this section a different solution is proposed based
	on: 1) computation of instantaneous estimates obtained by maximizing the IRF; 2) combination of the instantaneous estimates computed at different time instances.
	Hereafter, we discuss in some detail these two steps.
	
	\subsubsection{Computation of the instantaneous estimates}
	
	The instantaneous estimates  $\widehat \Theta _{t,\ell }^i$ and $\widehat \Gamma _{t,\ell }^i$ at time $t$ and consensus step $\ell$
	are computed by solving the optimization problem
	\begin{align}
	(\widehat \Theta _{t,\ell }^i , \widehat \Gamma _{t,\ell }^i) = 
	\mathop {\arg} \max \limits_{{\Theta ^i, \Gamma^i}} {\cal W}_{t,\ell}^i\left( {{\Theta ^i}} , \Gamma^i \right) \label{opt2}
	\end{align}
	To this end, a \textit{multiple initialization} strategy has to be adopted.
	Because of the large dimension of the parameter space and of the non-trivial dependence of the IRF on the drift parameters $\Gamma^i$,
	the choice of the initial points is a crucial issue. In what follows, we propose a procedure for the selection of the initial points
	that is based on some geometric insights on the sensor registration problem.
	
\begin{figure}[tb]
		\centering {
			\begin{tabular}{ccc}
				\includegraphics[width=0.45\textwidth]{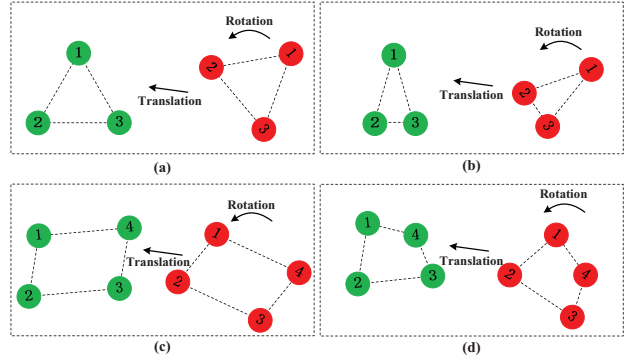}\\
			\end{tabular}
		}
		\caption{Example of sensor registration, where the circles in different colors represent the targets in the coordinates of two generic sensor nodes. 
			In the top subfigures three targets are located at the vertices: (a) of an equilateral triangle; (b) of an oblique triangle. 
			In the bottom subfigures, four targets are located at the vertices: (c) of a symmetrical quadrangle; 
			(d) of an asymmetrical quadrangle.}
		\vspace{-0.5\baselineskip}
		\label{Fig:TMSR}
	\end{figure}

	Generally speaking, in the context of multi-target tracking, sensor registration essentially amounts to matching the set of tracks of neighboring sensors 
	by \textit{rotation} and \textit{translation} operations (see Fig. \ref{Fig:TMSR}). With this respect, it is immediate to see that a necessary condition
	for sensor registration is that the number of tracks is at least equal to $3$ (in fact, there is not a unique way for matching points, i.e. a single track, or segments, i.e. two tracks).
	Recall now that, in the context of the GM-CPHD filter, each Gaussian component of the spatial density $s^i_{t,\ell-1}$ can be seen as a track for sensor $i$.
	Hence, each component $k \in \mathcal K^{i}_{t,\ell}$ in $ {\cal W}_{t,\ell}^i\left( {{\Theta ^i}} , \Gamma^i \right) $ can be seen as an association among the tracks $k(j)$, $j \in \mathcal N^i$, of the sensors in the neighborhood $\mathcal N^i$. 
	Then, by selecting a triplet $\{ k_1, k_2, k_3\} \subset \mathcal K^{i}_{t,\ell}$ of  components of $ {\cal W}_{t,\ell}^i\left( {{\Theta ^i}} , \Gamma^i \right) $,
	one has exactly three tracks $\{ k_1(j), k_2(j), k_3 (j)\}$ for any sensor $j \in \mathcal N^i$ so that a tentative matching 
	in terms of rotation $\Gamma _{\left\{ {{k_1},{k_2},{k_3}} \right \}}^i$ and translation  $\Theta _{\left\{ {{k_1},{k_2},{k_3}} \right\}}^i$
	can be easily computed (see Appendix B). Such tentative matchings can be used as initial points in the maximization (\ref{opt2}) 
	by applying the following steps: (i)
	list all the possible triplets of Gaussian components of the IRF; (ii)
	for each triplet $\left\{ {{k_1},{k_2},{k_3}} \right\}$, find the corresponding initial point $\left( \Theta _{\left\{ {{k_1},{k_2},{k_3}} \right\}}^i, \Gamma _{\left\{ {{k_1},{k_2},{k_3}} \right\}}^i \right)$ using the strategy described in Appendix B. 
	
	\subsubsection{Combination of the instantaneous estimates}
	
	Combination of the instantaneous estimates $(\widehat \Theta _{t,\ell }^i , \widehat \Gamma _{t,\ell }^i)$ computed at different $t$ and $\ell$ needs special care
	because some of such estimates can be unreliable under particular geometrical configurations of the targets present in the scenario.
	For instance, in subfigure (a), when the coordinates of node $j$ are rotated of $\gamma ^{i,j}$ or $\gamma ^{i,j}+120^o$ or $\gamma ^{i,j}+240^o$,
	the triangles can be perfectly matched.
	This can also happen when the rotation degree is $\gamma ^{i,j}$ or $\gamma ^{i,j}+180^o$ in subfigure (c).
	Hence, even when more than $3$ targets are present, there may exist unsolvable ambiguities between two sensor nodes.
	Such ambiguities always occur when:
	(a) the number of targets is odd and the targets are uniformly located on a circle;
	(b) the number of targets is even and the targets are located on the vertices of any symmetrical polygon.
	
	Generally speaking, when the moving targets are not organized as a group, 
	conditions (a) and (b) occur only occasionally,
	which is enough for sensor registration. However, the unreliable instantaneous estimates generated in the presence of such ambiguities need to be singled out and
	eliminated. For this reason, a simple averaging of the instantaneous estimates $(\widehat \Theta _{t,\ell }^i , \widehat \Gamma _{t,\ell }^i)$ is ruled out and, instead, 
	a multi-hypotheses approach is adopted wherein  a set of possible estimates 
	and their respective weights (defined in terms of reward functions) are maintained,
	i.e. $\Xi _{t,\ell }^i = \left\{ { {\widetilde \Theta _{t,\ell }^{i,h},\widetilde \Gamma _{t,\ell }^{i,h}} ,\kappa _{t,\ell }^{i,h}} \right\}_{k = 1}^{\widetilde{N}_{t,\ell }^{i}}$.
	At sampling time $t$ and consensus step $\ell$,
	the instantaneous estimates $\left( {\widehat \Theta _{t,\ell }^i,\widehat \Gamma _{t,\ell }^i} \right)$ are computed first,
	then followed by {\it association and estimation} as summarized in Table \ref{tab:SRBUDO},
	where $\delta _\theta$ and $\delta _\gamma$ are two preset thresholds.
	Note that one can limit the memory space by imposing a maximum number of elements in $\Xi _{t,\ell}^i$ so that,
	once $\widetilde{N}_{t,\ell }^{i}$ exceeds the limit,
	the element with minimal weight is deleted.

	\setcounter{magicrownumbers}{0}
	\begin{table} 
		\caption{Sensor registration with both unknown drift and orientation (node $i$, time $t$, consensus iteration $\ell$)}   \label{tab:SRBUDO}
		\begin{center}
			\begin{tabular}{l p{9.5cm}}
				\hline \hline	
				\textbf{Input:} & ${\Xi ^i_{t,\ell-1}}$ and $f_{t,\ell  - 1}^j\left( {\cal X}^j \right),j \in {{\cal N}^i}$  \\
				\hline \hline	
				\rownumber      & Find the instantaneous estimates $\widehat \Theta _{t,\ell }^i,\widehat \Gamma _{t,\ell }^i$ using a multi-initialization procedure  \\
				& \textbf{Association}  \\
				\rownumber      & Let $\mathcal{H} = \bigg \{ h: {{\left \| {\widehat \Theta _{t,\ell }^i - {{\widetilde \Theta }^{i,h}_{t,\ell-1}}} \right\|}_2} \le {\delta _\theta }$  $ \& \; {{\left\| 
						{\widehat \Gamma _{t,\ell }^i - {{\widetilde \Gamma }^{i,h}_{t,\ell-1}}} \right\|}_2} \le {\delta _\gamma } \bigg \}$  \\
				\rownumber      & If $ \mathcal{H} = \emptyset$  \\
				\rownumber      & $\;\;$ $\Xi _{t,\ell }^i = \Xi _{t,\ell  - 1}^i \bigcup \left( \widehat \Theta _{t,\ell }^i, \widehat \Gamma _{t,\ell }^i, \mathcal{W}_{t,\ell }^i 
				\left( \widehat \Theta _{t,\ell }^i, \widehat \Gamma _{t,\ell }^i \right) \right) $   \\
				\rownumber      & Else for $h \in \mathcal H$ \\
				\rownumber      & $\;\;$ Update the associated parameter set as  \\
				\rownumber      & $\;\;\;\;\;\;$ $\widetilde \Theta _{t,\ell }^{i,h} = \kappa^{i,h} \, \widehat \Theta _{t,\ell }^i + \left( {1 - {\kappa^{i,h}}} \right)\widetilde \Theta _{t,\ell  - 1}^{i,h}$ \\
				\rownumber      & $\;\;\;\;\;\;$ $\widetilde \Gamma _{t,\ell }^{i,h} = {\kappa^{i,h}} \, \widehat \Gamma _{t,\ell }^i + \left( {1 - {\kappa^{i,h}}} \right)\widetilde \Gamma _{t,\ell  - 1}^{i,h}$ \\
				\rownumber      & $\;\;$ where \\
				\rownumber      & $\;\;\;\;\;\;$ ${\kappa^{i,h}} = \frac{{{{\cal W}_{t,\ell }^i\left( {\widehat \Theta _{t,\ell }^i,\widehat \Gamma _{t,\ell }^i} \right)}}}{{{\kappa ^{i,h}_{t,\ell-1}} + {\cal W}_{t,\ell }^i\left( {\widehat \Theta _{t,\ell }^i,\widehat \Gamma _{t,\ell }^i} \right)}}$ \\
				\rownumber      & $\;\;$ Update the weight of the associated parameter as \\
				\rownumber      & $\;\;\;\;\;\;$ $\kappa _{t,\ell }^{i,h} = \kappa _{t,\ell  - 1}^{i,h} + \mathcal{W}_{t,\ell }^i\left( {\widehat \Theta _{t,\ell }^i,\widehat \Gamma _{t,\ell }^i} \right)$ \\
				\rownumber      & End \\
				& \textbf{Registration parameter estimation}  \\
				\rownumber      & Let $h^* = \mathop {\max }\limits_h \left\{ {\kappa _{t,\ell }^{i,h},h \in \left\{ {1, \dots, \widetilde{N}_{t,\ell }^{i}} \right\}} \right\}$ \\	
				\rownumber      & Then $\left( {\widehat \Theta _{1:t,\ell }^i,\widehat \Gamma _{1:t,\ell }^i} \right) = \left( {\widetilde \Theta _{t,\ell }^{i,h^*},\widetilde \Gamma _{t,\ell }^{i,h^*}} \right)$ \\		
				\hline \hline
			\end{tabular}
		\end{center}
	\end{table}

	\subsection{Distributed joint sensor registration and multitarget tracking}
	
	Combining the proposed sensor registration algorithm
	with the CCPHD filter, the algorithm of Table \ref{tab:JSSLCCPHD} for joint distributed sensor registration and multitarget tracking is obtained
	(for the sake of brevity, only the more general case of both unknown drift and orientation parameters is provided).
	Notice that, in practice,
	it may not be necessary/desirable to perform both sensor registration and consensus at all time instants.
	Specifically, the following practical suggestions can be given.
	\begin{itemize}
		\item It is better not to carry out consensus steps at the beginning when sensor registration has not yet been achieved,
		since the information provided by the TC may not be sufficient to provide a reliable estimate of $\Theta ^i$ and $\Gamma ^i$,
		so that performing fusion with an imprecise sensor registration could lead to a performance deterioration as compared to the local
		CPHD filters. 
		Hence, in practice, it is better to activate consensus only when a sufficient amount of data has been collected so
		that the sensor registration is reliable enough (e.g, after sensor registration has been performed a certain number of times).	\\
		\item
		At each sampling time, the sensor registration algorithm can be executed only once (for instance only when $ \ell = 1$),
		since sensor registration can be computationally demanding (as it involves an optimization routine),
		and the information about the drift and orientation parameters is maximal at the first consensus step.
		In the case in which only the drift parameters are needed (since the orientation parameters are already known), 
		one can further save computations
		by performing optimization of the TC only once every several time intervals. \\
		\item
		Sensor registration can be performed only when sufficiently many targets are detected by the sensors (i.e., the cardinality estimation is above a certain threshold).
		For instance, a single target is enough for sensor registration with known orientation parameters and, otherwise, at least three targets are needed.
	\end{itemize}

	\setcounter{magicrownumbers}{0}
	\begin{table}
		\caption{Joint distributed sensor registration and multitarget tracking (node $i$, time $t$)}   \label{tab:JSSLCCPHD}
		\begin{center}
			\begin{tabular}{l p{8cm}}
				\hline \hline
				\text{Input:}  & $f_{t-1}^i\left( {{{\cal X}^i}} \right)$ and $\Xi ^i_{t-1,L}$  \\
				\hline \hline
				\rownumber$\;$ & Carry out steps 1-2 in Table \ref{tab:GCICPHD} \\
				\rownumber$\;$ & Set $\Xi_{t,0}^i = \Xi_{t,L-1}^i$ \\
				\rownumber$\;$ & For $\ell = 0,\ldots,L-1$ \\ 
				\rownumber$\;$ & $\;\;$ Carry out {\it steps 4-5} in Table \ref{tab:GCICPHD} \\			
				\rownumber$\;$ & $\;\;$ If sensor registration has to be performed \\ 
				\rownumber$\;$ & $\;\;\;\;$ Compute ${\left( {\widehat \Theta _{1:t,\ell+1}^i,\widehat \Gamma _{1:t,\ell+1}^i} \right)}$ using the algorithm of Table  
				\ref{tab:SRBUDO} \\ 
				\rownumber$\;$ & $\;\;$ End if \\ 
				\rownumber$\;$ & $\;\;$ If consensus has to be performed \\ 
				\rownumber$\;$ & $\;\;\;\;$ Set $\Theta ^i = \widehat \Theta _{1:t,\ell+1}^i$, $ \Gamma^i = \widehat \Gamma _{1:t,\ell+1}^i$  \\
				\rownumber$\;$ & $\;\;\;\;$ Carry out step 6 in Table \ref{tab:GCICPHD}  \\
				\rownumber$\;$ & $\;\;$ Else \\ 
				\rownumber$\;$ & $\;\;\;\;$ Set $f_{t,\ell+1 }^i\left( {{{\cal X}^i}} \right) = f_{t,\ell}^i\left( {{{\cal X}^i}} \right)$  \\
				\rownumber$\;$ & $\;\;$ End \\ 
				\rownumber$\;$ & End for \\ 
				\rownumber$\;$ & Carry out steps 8-9 in Table \ref{tab:GCICPHD}  \\
				\hline \hline
				\text{Output:} & $f_t^i\left( {{{\cal X}^i}} \right)$ and $\Xi ^i_{t,L}$ \\
				\hline \hline
			\end{tabular}
		\end{center}
	\end{table}

	
	\section{Simulation experiments}
	
	In this section, the performance of the proposed algorithm is evaluated 
	by carrying out simulations on a 2-dimensional (planar) DMT scenario. 
	The surveillance region is a square of $8000 \times 8000\left[ {{m^2}} \right]$ which contains $6$ targets. 
	The target motion model used in the filter is a {\em white noise acceleration model}  \cite{Li2004Survey} with standard deviation of the acceleration
	equal to $3 \, [m / s^2]$ and sampling interval equal to $1\left[ s \right]$.
	
	
	The algorithm is tested on a nonlinear sensor network. For each sensor $i$,
	the single-target likelihood has the form $\mathcal L^i_t (y^i,x^i) = \mathcal G (y^i; h^i(x^i),R)$
	where
	\begin{align}
	h^i\left( {{x^i}} \right) = \left[ {\begin{array}{*{20}{cc}}
		{\sqrt {{{\left( {{\xi ^i}} \right)}^2} + {{\left( {{\eta ^i}} \right)}^2}} } &
		{{\mathop{\rm atan}\nolimits} 2\left( {{\xi ^i},{\eta ^i}} \right)}
		\end{array}} \right]^\top
	\end{align}
	and ${R^i} = {\mathop{\rm diag}\nolimits} \left( {\sigma _r^2,\sigma _\beta ^2} \right)$, with
	${\sigma _r} = 2\left[ m \right]$ and
	${\sigma _\beta } = 0.1\left[ {^o} \right]$. 
	{
		In this case, the GM representation of the spatial PDF (\ref{eq:GM})
		at each sensor node is propagated by employing the \emph{extended Kalman filtering} recursion,
		see (46)-(51) of \cite{vo2007analytic}.
	}
	
\begin{figure}[tb]
		\centering {
			\begin{tabular}{ccc}
				\includegraphics[width=0.45\textwidth]{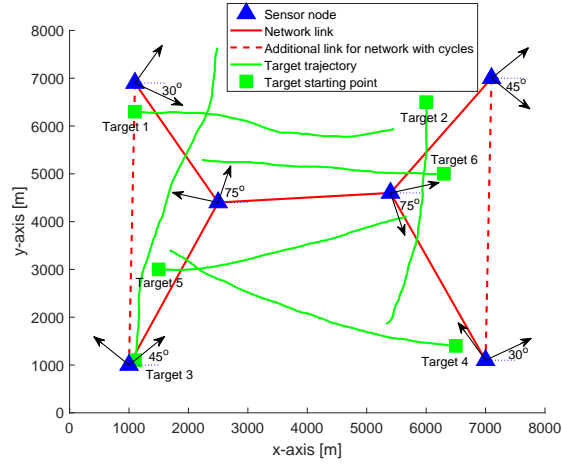}\\
			\end{tabular}
		}
		\caption{Considered simulation scenario: sensor networks and target trajectories.}
		\vspace{-0.5\baselineskip}
		\label{Fig:TYSN}
	\end{figure}

	The parameters of local CPHD filters are set as follows: $P_{s,t} = 0.9$, $P_{d,t}^i = 0.98$ for any $i \in {\cal N}$. 
	The maximum number of targets that the CPHD filters can handle is set to $N_{max} = 10$.
	For the target birth, we assume six high-likelihood zones that are known a priori. 
	Accordingly, at sensor node $i$, a $6$-component GM
	has been hypothesized for the birth intensity in the local coordinates of node $i$.
	The clutter set at each sensor node is assumed to be a Poisson point process with intensity $20$
	and uniform spatial distribution over the surveillance region.
	{The simulation horizon is set to $T=300[s]$ and the six targets appear/disappear as specified hereafter.
		Targets $1-4$ appear at $1[s]$,
		while targets $5$ and $6$ appear at $101 [s]$ and $121[s]$, respectively.
		Then target $1$ and $3$ disappear at $161[s]$ and $201[s]$ respectively, while the remaining targets ($2, 4, 5, 6$) stay in the surveillance region until the end of the simulation.}
	For each node $i$, the proposed algorithm starts only when the node itself and its neighbors detect more than 3 targets.
	Consensus is carried out starting from $t = 150[s]$ by using the estimated drift and orientation parameters to perform the
	changes of coordinates. 
	The number of consensus steps is set to $L=3$. 
	In order to save computational resources,
	after consensus begins, the sensor registration algorithm is run only when $\ell=1$ at each sensor node.
	For the approximation of the GM power, we adopted the same strategy as in \cite{battistelli2013consensus}. 
	The initial estimates of the drift parameters have been set to $\hat \theta _{0,0}^{i,j} = 0$ and $\hat \gamma _{0,0}^{i,j} = 0$, for any $\left( {i,j} \right) \in {\cal A}$.
	The proposed algorithm is tested on two different networks
	both consisting of 6 nodes: one with a tree topology and the other containing cycles.
	The considered simulation scenario is depicted in Fig. \ref{Fig:TYSN}.
	
\begin{figure}[tb]
		\centering {
			\begin{tabular}{ccc}
				\includegraphics[width=0.5\textwidth]{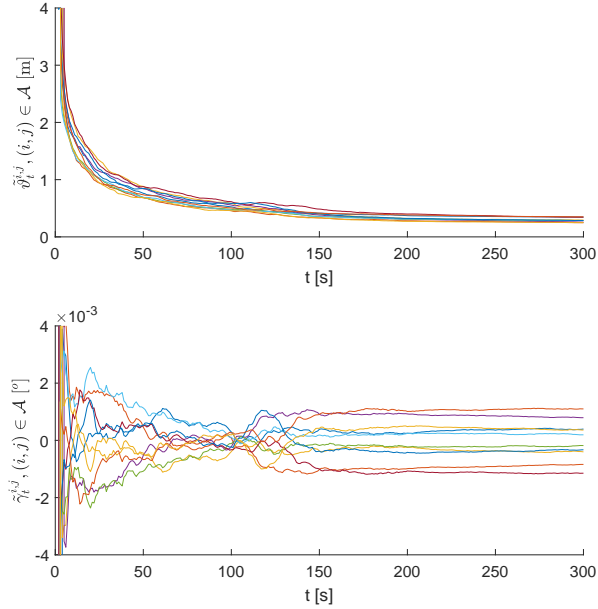}\\
			\end{tabular}
		}
		\caption{Performance of sensor registration in a  tree-network.
			The top and bottom subfigures provide the time-behavior of the registration parameter errors $\tilde \vartheta _{t}^{i,j}$ (drift) and, respectively,  $\tilde \gamma _t^{i,j}$ (orientation)
			for all $\left( {i,j} \right) \in {\cal A}$.}	 
		\vspace{-0.5\baselineskip}
		\label{Fig:PARtree}
	\end{figure}

\begin{figure}[tb]
		\centering {
			\begin{tabular}{ccc}
				\includegraphics[width=0.5\textwidth]{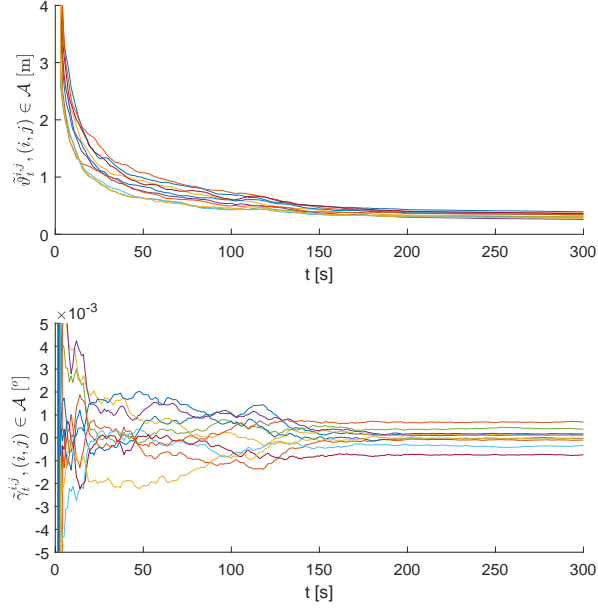}\\
			\end{tabular}
		}
		\caption{Performance of sensor registration in a  network with cycles. 
			The top and bottom subfigures provide the time-behavior of the registration parameter errors $\tilde \vartheta _{t}^{i,j}$ (drift) and, respectively,  $\tilde \gamma _t^{i,j}$ (orientation)
			for all $\left( {i,j} \right) \in {\cal A}$.}
		\vspace{-0.5\baselineskip}
		\label{Fig:PARcircle}
	\end{figure}

	Figs. \ref{Fig:PARtree}-\ref{Fig:PARcircle} analyze the sensor registration performance by displaying the time evolution,
	averaged over 200 Monte Carlo runs,
	of the drift parameter estimation errors
	{
		$ \tilde \vartheta  _{t}^{i,j}  = \sqrt {{{\left( {{\vartheta  ^{i,j}} - \hat \vartheta  _t^{i,j}} \right)}^\top}\left( {{\vartheta  ^{i,j}} - \hat \vartheta  _t^{i,j}} \right)} $ and the orientation parameter estimation errors
		$
		\tilde \gamma _{t }^{i,j} = {\gamma ^{i,j}} - \hat \gamma _{t }^{i,j}
		$
	}
	for the two networks of Fig. \ref{Fig:TYSN}.
	It can be seen that in all scenarios the estimated errors of drift and orientation parameters exhibit a stable behavior with satisfactory performance.
	
\begin{figure}[tb]
		\centering {
			\begin{tabular}{ccc}
				\includegraphics[width=0.45\textwidth]{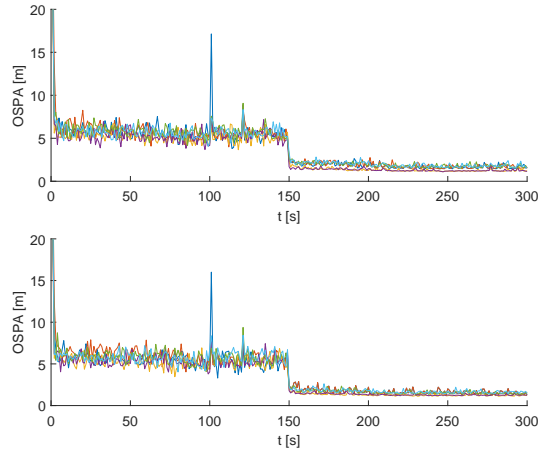}\\
			\end{tabular}
		}
		\caption{Time-behavior of the OSPA in each sensor node: for the network wit tree topology (top subfigure) and the network with cycles (bottom subfigure).}
		\vspace{-0.5\baselineskip}
		\label{Fig:OSPAall}
	\end{figure}

	Further, Fig. \ref{Fig:OSPAall} plots the time evolution,
	averaged over 200 Monte Carlo runs,
	of the \textit{optimal subpattern assignment} (OSPA) distance \cite{schuhmacher2008consistent} 
	(with order $p=2$ and cutoff $c=50\left[m\right]$) 
	between the targets and the estimated i.i.d. cluster RFS in each network node.
	It can be seen that when sensor registration has been achieved and the consensus algorithm begins to work ($t \ge 150s$),
	the performance of the DMT algorithm is much better as compared to the case without consensus ($t<150s$).
	
\begin{figure}[tb]
		\centering {
			\begin{tabular}{ccc}
				\includegraphics[width=0.5\textwidth]{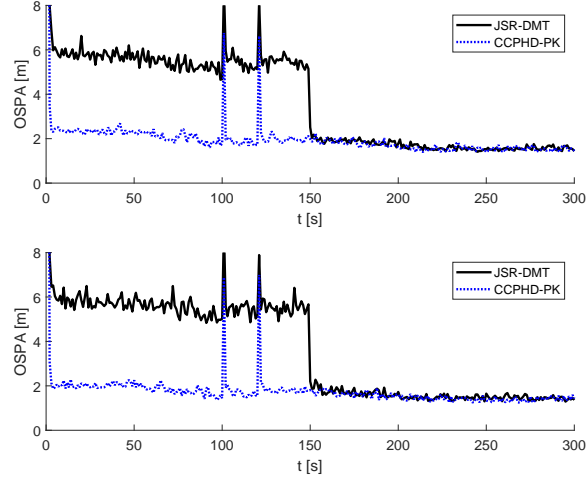}\\
			\end{tabular}
		}
		\caption{Time-behavior of the OSPA, averaged over all sensor nodes, with JSR-DMT (proposed method for joint sensor registration and DMT) 
			and CCPHD-PK (DMT with perfect knowledge of the registration parameters): for the network with tree topology (top subfigure) and the network with cycles (bottom subfigure).}
		\vspace{-0.5\baselineskip}
		\label{Fig:OSPACOM}
	\end{figure}

	{
		Finally, 
		Fig. \ref{Fig:OSPACOM} compares the OSPA of the proposed \textit{joint sensor registration}
		and  DMT algorithm (referred to as JSR-DMT) with 
		the one achievable by a CCPHD filter with perfect knowledge of the registration parameters  (referred to as CCPHD-PK) in the two cases of network with tree topology and network with cycles.
		It can be seen that once consensus starts ($t \ge 150 [s]$)
		the JSR-DMT exhibits almost the same accuracy as CCPHD-PK,
		thus demonstrating the effectiveness of the proposed sensor registration approach.}
	
	\section{Conclusions}
	The paper has dealt with distributed multitarget tracking over an unregistered sensor network.
	It has been shown how it is possible to jointly estimate, in a fully distributed and computationally feasible way, both the registration parameters (i.e. relative positions and orientations of the sensor nodes)
	as well as the number and kinematic states of the targets present in the surveillance region.
	The problem, referred to as distributed joint sensor registration and multitarget tracking, 
	has been solved by using a \textit{Cardinalized Probability Hypothesis Density} (CPHD) filter in each sensor node of the network,
	for the update of a local posterior multitarget density and then minimizing an information-theoretic criterion that measures the discrepancy among such local posteriors for both sensor fusion and registration purposes. 
	The effectiveness of the proposed approach has been successfully tested 
	via simulation experiments. 

	
	%

	\section*{Appendix A}
	
	{\em Proof of Proposition \ref{prop1}:} By recalling the definition of set integral, we have
	\begin{align}
	{\cal J}_{t,\ell }^i\left( \Theta ^i,\Gamma ^i \right) =  - \log \Bigg \{ \sum_{n=0}^\infty \frac{1}{n!} \int \prod\limits_{j \in {{\cal N}^i}} 
	\left[ f_{t,\ell  - 1}^{j,i} \left( \{ x_1^i , \ldots, x_n^i \} ;\theta^{i,j} , \gamma^{i,j} \right) \right]^{\omega^{i,j}}
	dx_1^i \cdots dx_n^i  \Bigg \}   \nonumber 
	\end{align}
	Since each $f_{t,\ell  - 1}^{j,i}$ is an i.i.d. cluster density, the above identity implies
	\begin{align}
	{\cal J}_{t,\ell }^i\left( \Theta ^i,\Gamma ^i \right) =  - \log \Bigg \{ \sum_{n=0}^\infty \frac{1}{n!} \int \prod\limits_{j \in {{\cal N}^i}} \Bigg ( \left[ n! p_{t,\ell  - 1}^{j} (n) \right]^{\omega^{i,j}} 
	\left[ \prod_{k=1}^{n} s_{t,\ell  - 1}^{j,i} \left( x_k^i;\theta^{i,j} , \gamma^{i,j} \right) \right]^{\omega^{i,j}} \Bigg )
	dx_1^i \cdots dx_n^i  \Bigg \}   \nonumber 
	\end{align}
	Recalling now that $\sum_{j \in \mathcal N^i} {\omega^{i,j}} = 1$ and, hence, $\prod\limits_{j \in {{\cal N}^i}} (n!)^{\omega^{i,j}}  = n!$, the above equation yields
	\begin{align}
	{\cal J}_{t,\ell }^i\left( \Theta ^i,\Gamma ^i \right) =  - \log \Bigg \{ \sum_{n=0}^\infty \prod\limits_{j \in {{\cal N}^i}} \left[ p_{t,\ell  - 1}^{j} (n) \right]^{\omega^{i,j}}  
	\Bigg ( \int \Bigg [ \prod_{j \in \mathcal N^i} s_{t,\ell  - 1}^{j,i} \left( x^i;\theta^{i,j} , \gamma^{i,j} \right) \Bigg]^{\omega^{i,j}}
	dx^i  \Bigg )^n \Bigg \}   \nonumber 
	\end{align}
	which can be rewritten as in (\ref{eq:CoCDGMCPHD}).
	\mbox{} \hfill $\square$ \vspace{.5cm}
	

	{\em Proof of Theorem \ref{prop:CALI}:}
	Observe preliminarily that the product of Gaussian distributions is again Gaussian and, specifically,
	\begin{align}
	\prod\limits_j {{\cal G}\left( {x;{\mu ^j},{P^j}} \right)}  = \det {\left( {2\pi \overline P} \right)^{1/2}}{\cal G}\left( {x;\overline \mu ,\overline P} \right)  \prod\limits_j {{\cal G}\left( {\overline \mu ;{\mu ^j},{P^j}} \right)}     \label{eq:POMG}
	\end{align}
	where $\overline P = {\left[ {\sum\nolimits_j {{{\left( {{P^j}} \right)}^{ - 1}}} } \right]^{ - 1}}$ and $\overline \mu  = \overline P ~\sum\nolimits_j {{{\left( {{P^j}} \right)}^{ - 1}}{\mu ^j}}$.
	Further, it is an easy matter to check that the following identity holds
	\begin{align}
	\prod\limits_j {{\cal G}\left( {\overline \mu ;{\mu ^j},{P^j}} \right)} = \left ( \prod_j \frac{1}{\det (2 \pi P^j)^{1/2}} \right )  \exp \left \{ -\frac{1}{2} \left [ \sum_j (\mu^j)^\top (P^j )^{-1} \mu^j - \overline \mu^\top \overline P^{-1}  \overline \mu\right ]\right \}
	\label{eq:POMG2}
	\end{align}
	
	Consider now the IRF (\ref{eq:IRF}). Substituting (\ref{eq:AOPGM}) and (\ref{eq:CoC}) into (\ref{eq:IRF}), we have
	\begin{align}
	{\cal W}_{t,\ell }^i\left( {{\Theta ^i},{\Gamma ^i}} \right) &= \int {\prod\limits_{j \in {{\cal N}^i}} {{{\left[ {s_{t,\ell  - 1}^{j,i}\left( {x;{\theta ^{i,j}},{\gamma ^{i,j}}} \right)} \right]}^{{\omega ^{i,j}}}}dx} } \nonumber  \\
	& = \int \prod\limits_{j \in {{\cal N}^i}} \sum_{k=1}^{N^j_{t,\ell-1}} \widehat \alpha _{t,\ell-1 }^{j,i,k} \, {\cal G}\left( {x; \widehat \mu _{t,\ell-1 }^{j,i,k}, \widehat P _{t,\ell-1 }^{j,i,k}} \right) d x
	\nonumber \\
	& = \sum_{k \in \mathcal K_{t,\ell}^i} \int  \prod\limits_{j \in {{\cal N}^i}} \widehat \alpha _{t,\ell-1 }^{j,i,k(j)} \, {\cal G}\left( {x; \widehat \mu _{t,\ell-1 }^{j,i,k(j)}, \widehat P _{t,\ell-1 }^{j,i,k(j)}} \right) d x
	\nonumber 
	\end{align}
	Hence, by employing (\ref{eq:POMG}), we get
	\begin{align}
	& {\cal W}_{t,\ell }^i\left( {{\Theta ^i},{\Gamma ^i}} \right) = \sum_{k \in \mathcal K_{t,\ell}^i} 
	\det {\left( {2\pi \overline P^{i,k}_{t,\ell}  (\Gamma^i)} \right)^{1/2}}  \int
	{  \cal G}\left( {x;\overline \mu^{i,k}_{t,\ell}  (\Gamma^i) , \overline P^{i,k}_{t,\ell}  (\Gamma^i)} \right) d x
	\prod\limits_j  \, \widehat \alpha _{t,\ell-1 }^{j,i,k(j)} \, {{\cal G}\left( {\overline \mu^{i,k}_{t,\ell}  (\Gamma^i) ; \widehat \mu _{t,\ell-1 }^{j,i,k(j)},{  \widehat P _{t,\ell-1 }^{j,i,k(j)}  }} \right)}   
	\nonumber
	\end{align}
	where
	$\overline P^{i,k}_{t,\ell}  (\Gamma^i) = {\left[ {\sum\nolimits_{j \in \mathcal N^i} {{{\left( {{ \widehat P_{t,\ell-1 }^{j,i,k(j)} }} \right)}^{ - 1}}} } \right]^{ - 1}}$ and $\overline \mu^{i,k}_{t,\ell}  (\Gamma^i)  = \overline P^{i,k}_{t,\ell}  (\Gamma^i)  ~\sum\nolimits_{j \in \mathcal N^i} {{{\left( {{ \widehat P _{t,\ell-1 }^{j,i,k(j)} }} \right)}^{ - 1}}{ \widehat \mu _{t,\ell-1 }^{j,i,k(j)}}}$.
	Noting that the integral in the above expression is equal to $1$ and exploiting the identity (\ref{eq:POMG2}), we obtain
	\begin{align}
	{\cal W}_{t,\ell }^i\left( {{\Theta ^i},{\Gamma ^i}} \right)  & = \sum_{k \in \mathcal K_{t,\ell}^i}  \det {\left( {2\pi \overline P^{i,k}_{t,\ell}  (\Gamma^i)} \right)^{1/2}} \times \left ( \prod_{j \in \mathcal N^i} \frac{\widehat \alpha _{t,\ell-1 }^{j,i,k(j)} }{\det (2 \pi   \widehat P_{t,\ell-1 }^{j,i,k(j)}    )^{1/2}} \right ) \nonumber \\
	& \quad \times  \exp \Bigg \{ -\frac{1}{2} \Bigg [ \sum_{j \in \mathcal N^i} \left (\widehat \mu _{t,\ell-1 }^{j,i,k(j)} \right )^\top \, \left ( \widehat P _{t,\ell-1 }^{j,i,k(j)}  \right )^{-1} \, \widehat \mu _{t,\ell-1 }^{j,i,k(j)}  - \left (\overline \mu^{i,k}_{t,\ell}  (\Gamma^i) \right )^\top \, \left (\overline P^{i,k}_{t,\ell}  (\Gamma^i) \right )^{-1}  \, \overline \mu^{i,k}_{t,\ell}  (\Gamma^i)  \Bigg ] \Bigg \} \label{eq:LCGM}
	\end{align}
	Observe now that, since the matrices $M^{i,j}$ are orthogonal, we have $\det (2 \pi   \widehat P_{t,\ell-1 }^{j,i,k(j)} ) = \det (2 \pi P_{t,\ell-1 }^{j,k(j)} / \omega^{i,j}) $.
	Further, recalling the definitions of $\Theta ^i$, ${\cal M}^i$, $\Psi _{t,\ell}^{i, k}$, ${E^i}$, and ${\bf{u}}_{t,\ell}^{i, k}$, it is
	immediate to check that $\overline P^{i,k}_{t,\ell}  (\Gamma^i) $ can be rewritten as in the statement of the theorem and, moreover, we can also write
	\begin{align}
	\overline \mu _{t,\ell }^{i, k} (\Gamma^i) = &  \overline P_{t,\ell }^{i, k}\left( {{\Gamma ^i}} \right) \bigg [ \left (E^i \right)^\top \left ( \Psi_{t,\ell}^{i,k} \right )^{-1} 
	\left ( \mathcal M^i {\bf u}_{t,\ell  - 1}^{i, k} + \Theta^i \right )
	+ \left ( P_{t,\ell-1}^{i,k(i)} / \omega^{i,i} \right )^{-1} \, \mu_{t,\ell-1}^{i,k(i)} \bigg ]
	\label{eq:FMV} 
	\end{align}
	\begin{align}
	&\sum_{j \in \mathcal N^i} \left (\widehat \mu _{t,\ell-1 }^{j,i,k(j)} \right )^\top \, \left ( \widehat P _{t,\ell-1 }^{j,i,k(j)}  \right )^{-1} \, \widehat \mu _{t,\ell-1 }^{j,i,k(j)}   \nonumber  \\
	& \quad = {\left( {{{\cal M}^i}{\bf{u}}_{t,\ell-1}^{i, k} + {\Theta ^i}} \right)^\top } \left ( \Psi _{t,\ell }^{i, k} \right )^{-1} \left( {{{\cal M}^i}{\bf{u}}_{t,\ell-1}^{i, k} + {\Theta ^i}} \right)  + {\left( { \mu _{t,\ell  - 1}^{i, k\left( i \right)}} \right)^\top } \left ( P_{t,\ell  - 1}^{i, k\left( i \right)} / \omega ^{i,i} \right )^{-1} \mu _{t,\ell  - 1}^{i,k\left( i \right)}   \label{eq:ET2RF}
	\end{align}
	Let us now define 
	\begin{align}
	{\cal E}_{t,\ell }^{i, k} \left( {{\Gamma ^i}} \right) &=  \left ( \Psi _{t,\ell }^{i, k} \right )^{-1} \nonumber  -  \left ( \Psi _{t,\ell }^{i, k} \right )^{-1} E^i 
	\overline P_{t,\ell}^{i,k}  (\Gamma^i) \left ( E^i \right )^\top  \left ( \Psi _{t,\ell }^{i, k} \right )^{-1}  \nonumber \\
	\varsigma _{t,\ell }^{i, k} \left( {{\Gamma ^i}} \right) &= \left ( \Psi _{t,\ell }^{i, k} \right )^{-1} E^i 
	\overline P_{t,\ell}^{i,k}  (\Gamma^i)
	{\left( {{{ P_{t,\ell  - 1}^{i, k\left( i \right)}}}/{\omega ^{i,i}}} \right)^{ - 1}} \mu _{t,\ell  - 1}^{i, k\left( i \right)} \nonumber
	\end{align}
	Then, by using the \emph{matrix inversion lemma}, we have
	\begin{align}
	{\left[ {{\cal E}_{t,\ell }^{i, k}\left( {{\Gamma ^i}} \right)} \right]^{ - 1}}  
	& = \Psi _{t,\ell }^{i, k} + {E^i}{\left[ {{{\left( { \overline P_{t,\ell }^{i,k} (\Gamma^i)} \right)}^{ - 1}} - {{\left( {{E^i}} \right)}^\top } \left ( \Psi _{t,\ell }^{i, k} \right)^{-1}{E^i}} \right]^{ - 1}}{\left( {{E^i}} \right)^\top }  \nonumber \\
	& =  {\Psi _{t,\ell }^{i, k}}  + {E^i}{{ P_{t,\ell  - 1}^{i, k\left( i \right)}}}{\left( {{E^i}} \right)^\top } / \omega ^{i,i}   = \Upsilon _{t,\ell }^{i, k}\left( {{\Gamma ^i}} \right) \nonumber 
	\end{align}
	\begin{align}
	{\left[ {{\cal E}_{t,\ell }^{i, k}\left( {{\Gamma ^i}} \right)} \right]^{ - 1}}\varsigma _{t,\ell }^{i, k}\left( {{\Gamma ^i}} \right)  
	& = E^i \overline P_{t,\ell}^{i,k}  (\Gamma^i)
	{\left( {{{ P_{t,\ell  - 1}^{i, k\left( i \right)}}}/{\omega ^{i,i}}} \right)^{ - 1}} \mu _{t,\ell  - 1}^{i, k\left( i \right)}   + {E^i}\left[ {{{ P_{t,\ell  - 1}^{i, k\left( i \right)}}}/{{{\omega ^{i,i}}}} -  \overline P_{t,\ell }^{i, k}\left( {{\Gamma ^i}} \right)} \right]{\left( {{{ P_{t,\ell  - 1}^{i, k\left( i \right)}}}/{{{\omega ^{i,i}}}}} \right)^{ - 1}} \mu _{t,\ell  - 1}^{i, k\left( i \right)}  \nonumber  \\
	& = {E^i} \mu _{t,\ell  - 1}^{i, k\left( i \right)} \nonumber
	\end{align}
	\begin{align}
	& {\left[ {\varsigma _{t,\ell }^{i, k}\left( {{\Gamma ^i}} \right)} \right]^\top }{\left[ {{\cal E}_{t,\ell }^{i, k}\left( {{\Gamma ^i}} \right)} \right]^{ - 1}}\varsigma _{t,\ell }^{i, k}\left( {{\Gamma ^i}} \right) \nonumber  \\
	& = {\left( { \mu _{t,\ell  - 1}^{i, k\left( i \right)}} \right)^\top }{\left( {{{ P_{t,\ell  - 1}^{i, k\left( i \right)}}}/{{{\omega ^{i,i}}}}} \right)^{ - 1}} \overline P_{t,\ell }^{i, k}\left( {{\Gamma ^i}} \right)  \left\{ {{{\left[ { \overline P_{t,\ell }^{i, k}\left( {{\Gamma ^i}} \right)} \right]}^{ - 1}} - {{\left( {{{ P_{t,\ell  - 1}^{i, k\left( i \right)}}}/{{{\omega ^{i,i}}}}} \right)}^{ - 1}}} \right\} \mu _{t,\ell  - 1}^{i, k\left( i \right)}   \nonumber  \\
	& = {\left( { \mu _{t,\ell  - 1}^{i, k\left( i \right)}} \right)^\top }{\left( {{{ P_{t,\ell  - 1}^{i, k\left( i \right)}}}/{{{\omega ^{i,i}}}}} \right)^{ - 1}} \mu _{t,\ell  - 1}^{i, k\left( i \right)} - {\left( { \mu _{t,\ell  - 1}^{i, k\left( i \right)}} \right)^\top }{\left( {{{ P_{t,\ell  - 1}^{i, k\left( i \right)}}}/{{{\omega ^{i,i}}}}} \right)^{ - 1}} \overline P_{t,\ell }^{i, k}\left( {{\Gamma ^i}} \right){\left( {{{ P_{t,\ell  - 1}^{i, k\left( i \right)}}}/{{{\omega ^{i,i}}}}} \right)^{ - 1}} \mu _{t,\ell  - 1}^{i, k\left( i \right)} 
	\nonumber
	\end{align}
	Considering (\ref{eq:FMV})-(\ref{eq:ET2RF}) as well as the above identities,
	the exponential argument in (\ref{eq:LCGM}) can be rewritten as in (\ref{eq:LCGM2}).
	Finally, by substituting (\ref{eq:LCGM2}) into (\ref{eq:LCGM}), the proof is concluded.
	\mbox{} \hfill $\square$
	
	\begin{align}
	&  \sum_{j \in \mathcal N^i} \left (\widehat \mu _{t,\ell-1 }^{j,i,k(j)} \right )^\top \, \left ( \widehat P _{t,\ell-1 }^{j,i,k(j)}  \right )^{-1} \, \widehat \mu _{t,\ell-1 }^{j,i,k(j)} 
	- \left (\overline \mu^{i,k}_{t,\ell} (\Gamma^i) \right )^\top \, \left (\overline P^{i,k}_{t,\ell}  (\Gamma^i) \right )^{-1}  \, \overline \mu^{i,k}_{t,\ell}  (\Gamma^i)
	\nonumber   \\
	& = {\left( {{{\cal M}^i}{\bf{u}}_{t,\ell }^{i, k} + {\Theta ^i}} \right)^\top }{\cal E}_{t,\ell }^{i, k}\left( {{\Gamma ^i}} \right)\left( {{{\cal M}^i}{\bf{u}}_{t,\ell }^{i, k} + {\Theta ^i}} \right) - 2{\left( {{{\cal M}^i}{\bf{u}}_{t,\ell }^{i, k} + {\Theta ^i}} \right)^\top }\varsigma _{t,\ell }^{i, k} + {\left[ {\varsigma _{t,\ell }^{i, k}\left( {{\Gamma ^i}} \right)} \right]^\top }{\left[ {{\cal E}_{t,\ell }^{i, k}\left( {{\Gamma ^i}} \right)} \right]^{ - 1}}\varsigma _{t,\ell }^{i, k}\left( {{\Gamma ^i}} \right)   \nonumber  \\
	& =  {\left( {{{\cal M}^i}{\bf{u}}_{t,\ell }^{i, k} + {\Theta ^i}} - {\left[ {{\cal E}_{t,\ell }^{i, k}\left( {{\Gamma ^i}} \right)} \right]^{ - 1}}\varsigma _{t,\ell }^{i, k}\left( {{\Gamma ^i}} \right)  \right)^\top }
	{\cal E}_{t,\ell }^{i, k}\left( {{\Gamma ^i}} \right)
	\left( {{{\cal M}^i}{\bf{u}}_{t,\ell }^{i, k} + {\Theta ^i}} - {\left[ {{\cal E}_{t,\ell }^{i, k}\left( {{\Gamma ^i}} \right)} \right]^{ - 1}}\varsigma _{t,\ell }^{i, k}\left( {{\Gamma ^i}} \right) \right)   \nonumber  \\
	& = {\left[ {{\Theta ^i} - \left( {{E^i} \mu _{t,\ell  - 1}^{i,k\left( i \right)} - {{\cal M}^i}{\bf{u}}_{t,\ell }^{i, k}} \right)} \right]^\top }{\cal E}_{t,\ell }^{i, k}\left( {{\Gamma ^i}} \right)\left[ {{\Theta ^i} - \left( {{E^i} \mu _{t,\ell  - 1}^{i,k\left( i \right)} - {{\cal M}^i}{\bf{u}}_{t,\ell }^{i, k}} \right)} \right] 
	\label{eq:LCGM2}
	\end{align}

	\section*{Appendix B}   \label{SEC:ApB}
	
	Recall that the Gaussian components are defined with respect to both position and velocity. However, by construction, 
	$\Theta^i = col \left ( T \vartheta^{i,j} , j \in \mathcal N^i \setminus \{ i \}\right )$. Hence, we are interested in matching only the positions of the tracks.
	For a given component $k \in \mathcal K_{t,\ell}^i $, the positions of the tracks $k(j)$, $j \in \mathcal N^i$, can be matched by choosing $\Theta^i$ and $\Gamma^i$ such that
	\[
	(\mathcal T^i)^\top \Theta^i - (\mathcal T^i)^\top \phi_{t,\ell}^{i.k} (\Gamma^i) = 0 
	\]
	where $(\mathcal T^i)^\top \Theta^i = col \left ( \vartheta^{i,j} , j \in \mathcal N^i \setminus \{ i \}\right )$, and, in view of (\ref{eq:IRFME}), we have
	\begin{align}
	\phi _{t,\ell }^{i, k}\left( {{\Gamma ^i}} \right) = col\left( { \mu _{t,\ell  - 1}^{i,k\left( i \right)} - {M^{i,j}}\mu _{t,\ell  - 1}^{j,k\left( j \right)},j \in {{\cal N}^i}\backslash \left\{ i \right\}} \right) \, .
	\nonumber
	\end{align}
	By writing
	${\mu}_{t,\ell  - 1}^{j, k \left( j \right)} = \left[ {\xi}_{t,\ell  - 1}^{j,k\left( j \right)} \; {\dot \xi}_{t,\ell  - 1}^{j, k\left( j \right)} \;
	{\eta}_{t,\ell  - 1}^{j, k\left( j \right)} \; \dot {\eta}_{t,\ell  - 1}^{j, k\left( j \right)} \right]^\top , \; j \in \mathcal N^i ,$
	it is an easy matter to check that the following identity holds
	\begin{align}
	\left( {\cal T} ^i \right)^\top \phi _{t,\ell }^{i,k}\left( {{\Gamma ^i}} \right)   = col\left( {b_{t,\ell  - 1}^{i,k\left( i \right)} - A_{t,\ell  - 1}^{j,k\left( j \right)}{{\varpi ^{i,j}}},j \in {{\cal N}^i}\backslash \left\{ i \right\}} \right)   \nonumber
	\end{align}
	where
	\begin{align}
	{\varpi ^{i,j}} &= {\left[ {\cos {\gamma ^{i,j}} \; \sin {\gamma ^{i,j}}} \right]^\top }  \nonumber \\
	b_{t,\ell  - 1}^{i, k\left( i \right)} &= {\left[ { \xi _{t,\ell  - 1}^{i,k\left( i \right)} \; \eta _{t,\ell  - 1}^{i, k\left( i \right)}} \right]^\top }  \nonumber \\
	A_{t,\ell  - 1}^{j, k\left( j \right)} &= \left[ {\begin{array}{*{20}{c}}
		{ \xi _{t,\ell  - 1}^{j, k\left( j \right)}}&{ -  \eta _{t,\ell  - 1}^{j, k\left( j \right)}}\\
		{ \eta _{t,\ell  - 1}^{j, k\left( j \right)}}&{\xi _{t,\ell  - 1}^{j, k\left( j \right)}}
		\end{array}} \right] \nonumber
	\end{align}
	Consider now a triplet $\{k_1,k_2,k_3\} \subset \mathcal K^i_{t,\ell}$. Given ${\varpi ^{i,j}}$,
	the orientation parameter $\gamma ^{i,j}$ can be easily obtained. Then, the positions of three sets of tracks can be matched by choosing  $\Theta^i$ and $\Gamma^i$ such that
	\begin{equation}
	\left \{ 
	\begin{array}{rcl}
	(\mathcal T^i)^\top \Theta^i - (\mathcal T^i)^\top \phi_{t,\ell}^{i.k_1} (\Gamma^i) &\approx& 0 \\
	(\mathcal T^i)^\top \Theta^i - (\mathcal T^i)^\top \phi_{t,\ell}^{i.k_2} (\Gamma^i) &\approx& 0 \\
	(\mathcal T^i)^\top \Theta^i - (\mathcal T^i)^\top \phi_{t,\ell}^{i.k_3} (\Gamma^i) &\approx& 0 
	\end{array}
	\right . \label{eq:matching}
	\end{equation}
	where the approximation symbol accounts for the fact that, in general, the three conditions cannot be exactly satisfied together due to the uncertainties in the tracks. Then, 
	the initial point associated to the triplet $\{k_1,k_2,k_3\}$ can be obtained by finding a solution 
	of (\ref{eq:matching}), in the least-squares sense, by means of the simple procedure of Table \ref{tab:IPFDO}.

	\setcounter{magicrownumbers}{0}
	\begin{table} 
		\caption{Find the initial point of the drift and orientation parameters for triplet $\left\{ {{k_1},{k_2},{k_3}} \right\}$ (node $i$, time $t$, consensus iteration $\ell$)}  
		\label{tab:IPFDO}
		\begin{center}
			\begin{tabular}{l p{7cm}}
				\hline \hline	
				\rownumber & For ${j \in {{\cal N}^i}\backslash \left\{ i \right\}}$:  \\
				\rownumber & $\;\;$ Define: \\
				\rownumber & $\;\;\;\;$ $A_{t,\ell  - 1}^{i,j} = \left[ {\begin{array}{*{20}{c}}
					{A_{t,\ell  - 1}^{j,{k_1}\left( j \right)} - A_{t,\ell  - 1}^{j,{k_2}\left( j \right)}}\\
					{A_{t,\ell  - 1}^{j,{k_1}\left( j \right)} - A_{t,\ell  - 1}^{j,{k_3}\left( j \right)}}
					\end{array}} \right]$  \\
				\rownumber & $\;\;\;\;$ $b_{t,\ell  - 1}^{i,j} = \left[ {\begin{array}{*{20}{c}}
					{b_{t,\ell  - 1}^{i,{k_2}\left( i \right)} - b_{t,\ell  - 1}^{i,{k_1}\left( i \right)}}\\
					{b_{t,\ell  - 1}^{i,{k_3}\left( i \right)} -  b_{t,\ell  - 1}^{i,{k_1}\left( i \right)}}
					\end{array}} \right]$  \\
				\rownumber & $\;\;\;\;$ Find ${\varpi ^{i,j}}$ by:  \\
				\rownumber & $\;\;\;\;\;\;\;\;$ $\left\{ \begin{array}{l}
				\mathop {\min }\limits_{{\varpi ^{i,j}}} {\left\| {A_{t,\ell  - 1}^{i,j}{\varpi ^{i,j}} - b_{t,\ell  - 1}^{i,j}} \right\|_2}\\
				s.t. \;\; {\left\| {{\varpi ^{i,j}}} \right\|_2} = 1
				\end{array} \right.$ \\
				\rownumber & $\;\;\;\;$ Let $\gamma _{ini}^{i,j} = {\mathop{\rm atan}\nolimits} 2 \left [ \varpi^{i,j} (1) , \varpi^{i,j} (2) \right ]$ \\
				\rownumber & $\;\;\;\;$ Let $\vartheta _{ini}^{i,j} = \frac{1}{3}\sum\nolimits_{m = 1}^3 {\left[ {b_{t,\ell  - 1}^{i,{k_m}\left( i \right)} - A_{t,\ell  - 1}^{j,{k_m}\left( j \right)}{\varpi ^{i,j}}} \right]} $ \\
				\rownumber & End for\\
				\rownumber & Set $\Theta _{\left\{ {{k_1},{k_2},{k_3}} \right\}}^i = col\left( {T\vartheta _{ini}^{i,j},j \in {{\cal N}^i}\backslash \left\{ i \right\}} \right)$\\
				\rownumber & Set $\Gamma _{\left\{ {{k_1},{k_2},{k_3}} \right\}}^i = col\left( {\gamma _{ini}^{i,j},j \in {{\cal N}^i}\backslash \left\{ i \right\}} \right)$\\
				\hline \hline
			\end{tabular}
		\end{center}
	\end{table}
	


	\ifCLASSOPTIONcaptionsoff
	\newpage
	\fi

	
	
	%
	
	\bibliographystyle{IEEEtran}
	\bibliography{SSLCPHD}
\end{document}